\documentclass{aa}  

\usepackage{graphicx}
\usepackage{txfonts}
\usepackage{hyperref}
\usepackage{subcaption}
\usepackage{color}
\usepackage{xcolor}
\usepackage{hyperref}
\usepackage{ulem}

\begin{document}

   \title{Traces of the evolution of cosmic void galaxies: An Integral Field Spectroscopy based analysis}

   \author{Agust\'in M. Rodr\'iguez-Medrano
          \inst{1,2} \fnmsep\thanks{Email: arodriguez@unc.edu.ar}
          \and
          Dante J. Paz  \inst{1,2}
          \and
          Dami\'an Mast \inst{2,3} \and \\
          Federico A. Stasyszyn \inst{1,2} 
          \and
          Andr\'es N. Ruiz \inst{1,2}
          }

   \institute{Instituto de Astronom\'ia te\'orica y experimental-Conicet, Laprida 854, Córdoba, Argentina\\
    \and
             Observatorio Astronómico de Córdoba, Universidad Nacional de Córdoba, Laprida 854, X5000BGR Córdoba, Argentina\\
    \and 
    Consejo de Investigaciones Científicas y Técnicas de la República Argentina, Buenos Aires, Argentina \\
                          }

   \date{Received November 21, 2024; accepted June 12, 2025}

 
  \abstract
   {Galaxies in the most underdense regions of the Universe, known as cosmic voids, exhibit astrophysical properties that suggest a distinct evolutionary path compared to galaxies in denser environments. Numerical simulations indicate that the assembly of void galaxies occurs later, leading to galaxies with younger stellar populations, low metallicities, and a high gas content in their halos, which provides the fuel to sustain elevated star formation activity.}
   {Our objective in this work is to test these numerical predictions with observational data by comparing galaxies in voids with galaxies in non-void environments.  }
   { We used voids identified in SDSS data and galaxies from the MaNGA survey, which provides galaxies with integral field spectroscopy (IFS). This IFS allows for state-of-the-art modeling of their stellar populations. We separated the galaxies into void and non-void samples, mimicked the magnitude distribution, and compared their integrated astrophysical properties as well as the metallicity and age profiles through a stacking technique, analyzing early type (ETGs) and late type (LTGs) galaxies separately.}
   { We find that void galaxies tend to have younger and less metal-rich stellar populations. This trend is observed both as a function of mass and in samples with matched magnitude distributions. Regarding gas mass, we do not find differences across environments. When dividing galaxies into ETGs and LTGs, we observe that ETGs show negative gradients in both age and metallicity, with void galaxies consistently appearing younger and less metal-rich. For LTGs, age gradients are also negative, showing younger populations in void galaxies. However, we do not find statistically significant differences in stellar metallicity gradients between void and non-void environments.
   }
   {Our results show how the astrophysical properties of galaxies in voids differ from those of galaxies in the rest of the Universe. This suggests that the void environment plays a role in the evolution of its galaxies, delaying their assembly and growth.}

\authorrunning{Rodr\'iguez-Medrano et al.}
\titlerunning{Traces of the evolution of void galaxies}
   \maketitle
%
\section{Introduction}
Galaxy evolution is driven by both in-situ processes and environmental effects. The large-scale and local densities of the galactic environment influence the astrophysical properties of galaxies. In this context, the most underdense regions of the universe, known as cosmic voids, are particularly interesting for studying galaxies. In these regions, galaxies evolve in an environment that can be free from interactions between them. As a result, these galaxies may have populations more representative of in-situ processes \citep{VanDeWeygaert2011, Kreckel2011}.

Other authors have previously studied galaxies in voids using observational data. In general, studies show that these environments, in contrast to galaxies in denser regions, host a population of bluer galaxies \citep{Rojas2004, vonBenda2008, Hoyle2012}, with higher star formation rates \citep{Rojas2005, Ceccarelli2008, Moorman2016}, and typically of late-type morphology \citep{Hoyle2005, Porter2023, ArgudoFernandez2024}. All these properties are consistent with the idea of delayed evolution for galaxies in such environments. In \cite{Dominguez2023b}, a sample of spectra from galaxies in voids and denser environments was used to reconstruct the star formation histories of these galaxies, revealing that those in voids exhibit delayed star formation histories compared to other samples.

Another approach to studying the influence of the environment is through numerical simulations. Modern simulations generally produce a population of void galaxies with properties qualitatively comparable to those in observational data \citep{RosasGuevara2022, Curtis2024}. One advantage of numerical simulations is their ability to trace the assembly history of galaxies through merger trees. These studies also confirm a delayed evolution for galaxies in voids compared to those in denser environments \citep{Martizzi2020, Alfaro2020, RodriguezMedrano2022, RodriguezMedrano2024}.

Differences in galaxy assembly translate into variations in the ages of their stellar populations, which can be reflected in their metallicities. In \citet{RodriguezMedrano2024}, using the IllustrisTNG simulation \citep{Nelson2019a, Pillepich2018, Springel2018}, we found lower stellar metallicities, at a given galaxy mass, for galaxies in voids compared to the general population. This is consistent with results from observational data \citep{Dominguez2023a} based on a sample of galaxies from SDSS-DR7 \citep{Abazajian2009}, with optical spectra integrated over the central regions of each galaxy (3 arcsec aperture).

Another feature detected in void environments is a lower stellar-to-halo mass relation (SHMR) for the galaxies \citep{Habouzit2020, RosasGuevara2022}. At a given stellar mass, void galaxies tend to have larger dark matter halos, which can retain more gas \citep{RodriguezMedrano2024}. 
As shown in \citet{ScholtDiaz2023}, halo mass influences the star formation history of galaxies. Galaxies with younger stellar populations inhabit halos with a lower stellar-to-halo mass ratio \citep{ScholtzDiaz2022}.
These predictions of the SHMR in numerical simulations have also been explored in observational data. The stellar-to-halo mass relation in void galaxies was investigated by \citet{Douglass2019}, who found no significant difference between the halo masses in void and non-void environments. We also studied this relation in \citep{RodriguezMedrano2023} using SDSS galaxies with halo masses obtained through the method of \citep{Rodriguez2020} but did not obtain conclusive results. Regarding gas content, \citep{Florez2021} detected an excess of gas in void galaxies. Therefore, the question of halo mass, and particularly the gas content of galaxies in voids, remains an open topic.

Integral field spectroscopy (IFS) is an observational technique that captures spectral information across a two-dimensional (2D) field of view. This provides two spatial dimensions plus an additional spectral dimension, forming what is known as a datacube. IFS has proven to be very effective for studying the stellar composition of galaxies with spatial resolution, capturing variations in stellar populations and other properties across galactic structures \citep{Bacon01,Sanchez2012AA,Croom2012MNRAS}.
This technique is used, for example, in the CAVITY project \citep{Perez2024} to study galaxies in voids. The sample of void galaxies from this project exhibits lower stellar mass surface density, younger ages, and higher specific star formation rates \citep{Sanchez2024}. These differences in the properties of void galaxies are present in all morphological types\citep{Conrado2024}.

The Mapping Nearby Galaxies at APO (MaNGA) survey, part of the SDSS-IV project, is currently the largest integral field spectroscopy survey \citep{Bundy2015}. MaNGA has observed approximately 10,000 galaxies in the local universe ($z<0.15$). These data have been used to employ methods such as the “fossil record method”, which leverages spatial resolution to apply spectral energy distribution (SED) models and deduce the star formation histories (SFH) of galaxies, and chemical enrichment histories (ChEH) \citep{IbarraMedel2016,Sanchez2022,Riffel2023}. This extensive sample enables studies on the relationship between galaxy evolution and their cosmic environment, particularly in cosmic voids. Since MaNGA galaxies are part of the original Sloan Digital Sky Survey (SDSS) sample \citep{York2000}, we can assign MaNGA galaxies to voids identified in a complete volume-limited sample in the SDSS survey. This allows us to compare galaxies in voids with those in denser environments and to investigate evolutionary differences and astrophysical properties based on environment, thereby mapping the impact of large-scale structures on galaxies.

This paper is organized as follows: in Section \ref{sec:data}, we describe the data used in this work, including the galaxy sample and the void catalog. In Section \ref{sec:results}, we present our results, dividing the section into different parts that examine the integrated properties of galaxies, the results related to galaxy morphologies, and the galaxy gradients. In Section \ref{sec:discusions}, we discuss the implications of our analysis, and in Section \ref{sec:conclusions}, we summarize the main conclusions of this study.
   
\section{Data}
\label{sec:data}

The MaNGA \citep{Bundy2015} and SDSS \citep{Bacon01} surveys occupy the same spatial region, ensuring substantial overlap that allows for effective integration of data from both sources. This overlap is advantageous as it enables the use of voids identified within SDSS, which provides a dense sample of galaxies suitable for robust identification of underdense regions in the universe. Subsequently, we study galaxies from the MaNGA survey, utilizing its integral field spectroscopy (IFS) capabilities, which are particularly beneficial for efficiently modeling stellar populations and their properties in general. IFS offers comprehensive spatial coverage by capturing spectral information across a two-dimensional field, allowing for detailed analysis of various stellar properties at different locations within galaxies. This spatially resolved spectroscopy makes it possible to examine gradients and variations in key stellar attributes, providing a more holistic view of galactic evolution. 
The MaNGA survey is designed to obtain integral field spectroscopy of nearby galaxies, and its main sample is divided into two subsamples: the Primary Sample and the Secondary Sample. These subsamples are optimized to cover different regions of galaxies: the Primary Sample includes galaxies with coverage extending at least 
$1.5\,R_e$ (effective radius), enabling detailed studies of their inner regions, while the Secondary Sample extends this coverage at least $2.5\,R_e$, facilitating the analysis of their outer parts, such as stellar halos and large-scale dynamics.

As described in \citep{Yan2016} and \citep{Wake2017}, due to this difference in spatial coverage, the secondary sample consists of galaxies observed at higher redshifts compared to the primary sample. This arises because observing more distant galaxies is necessary to cover a larger area with the same IFU size.

For the primary sample, the median signal-to-noise ratio (S/N) per fiber in the outer regions at $1.5\,R_e$ is $8.3$, reaching a S/N of $37.3$ in an elliptical annulus after stacking the fibers. For the secondary sample, the median S/N per fiber in the outer regions (in this case at $2.5\,R_e$) is $2.3$, increasing to $11.4$ in an elliptical annulus after stacking.

By leveraging the overlap between SDSS and MaNGA, this study aims to shed light on the evolutionary differences that arise from environmental variations within large cosmic structures.

\subsection{MaNGA: Average properties and morphology}
\label{sec:data_integradas}

In this work, we used the dataproducts extracted from the datacubes with the pyPipe3D pipeline, applied to MaNGA data from SDSS-DR17 \citep{Sanchez2016bRMxAA,Sanchez2016aRMxAA,Sanchez2022, Lacerna2022}. This analysis includes both integrated and spatially resolved properties of the stellar population and the ionized gas.
The pyPipe3D code, employs a binning method based on continuum segmentation to enhance the signal-to-noise ratio while preserving the spatial structure of galaxies. This analysis utilizes the \textsc{MaStar\_sLOG} stellar library \citep{Yan2019}, which includes simple stellar populations (SSPs) with 39 ages and 7 metallicities, along with a Salpeter initial mass function (IMF). Additionally, it incorporates kinematic parameters such as stellar velocity and velocity dispersion, treated as free parameters, while accounting for interstellar dust extinction effects.

In this work, we use only those galaxies that have a high signal-to-noise ratio, are not merging systems, do not have issues in redshift determination, or are not contaminated by a bright star in the field of view. All of these criteria are indicated by the QCFLAG of the catalogue used.

We use the second version of the morphological classification obtained by \cite{Vazquez-Mata2022} from the visual classification of MaNGA galaxies in SDSS-DR17, using mosaics generated by combining $r$-band images from SDSS and the DESI Legacy Survey \citep{DeyDESI2019}.

\subsection{MaNGA: Stellar profiles}
\label{sec:data_gradients}
To analyse the radial behaviour of certain properties in our sample, 
we used radial profiles constructed from the publicly available MEGACUBES dataset\footnote{\url{http://www.if.ufrgs.br/\~riffel/software/megacubes/}} \citep{Riffel2023}. These datacubes provide maps of various properties and emission-line profiles for each spaxel, along with a table of average properties over different galaxy radii. For the stellar populations, measurements have been made using full spectral fitting for stellar population synthesis on the datacubes, using the STARLIGHT code \citep{CidFernandes2005,CidFernandes2018}. In addition, these MEGACUBES contain absorption-free emission line datacubes. Emission line fitting was performed using the IFSCUBE3 Python package \citet{Ruschel-Dutra2020, Ruschel-Dutra2021} 
to fit the profiles of the most prominent optical emission lines. 

The analysis employs a binning method that enhances the signal-to-noise ratio while preserving spatial resolution. The stellar libraries used are based on updated MILES models \citep{Gonzalez-Delgado2005, Vazdekis2010,  Vazdekis2016}, covering a range of 21 ages and 4 metallicities. Additionally, a Salpeter initial mass function is adopted. The fits include kinematic parameters, such as stellar velocity and velocity dispersion, which are incorporated as key elements in the spectral modelling.

The profiles provided by MEGACUBES extend up to at least $1.5\,R_e$  or $2\,R_e$, depending on whether the galaxies belong to the Primary or Secondary sample, respectively.
The catalogue presents metallicity and age profiles for each galaxy, where the profiles are computed by averaging the spaxels that fall within specific radial intervals, expressed in units of the effective radius  ($R_e$). 

\subsection{Void and non-void galaxies}

We use the void finder algorithm described in \citet{Ruiz2015, Ruiz2019} to identify spherical voids in a sample of galaxies of the SDSS-DR12 catalogue \citet{Alam2015}. Voids were identified by finding spheres in a galaxy distribution with integrated contrast density ($\Delta$) below some threshold, in our case we use $\Delta<-0.9$. The radius of the void ($r_{\rm void}$) is defined by the distance, from the centre of the sphere, at which a $\Delta=-0.9$ is reached.
The tracer galaxies selected are at $z>0.03$, have an absolute magnitude in r-band $M_r< -20.19$ and constitute a completed volume sample at $z<0.12$. We assume a cosmology with $\Omega_M = 0.31$ and $\Omega_{\Lambda} = 0.69$. We selected all voids with $r_{\rm void}>12 h^{-1}$Mpc to avoid spurious voids due to shot noise.

For this selection of voids, we computed the distance from each galaxy in the MaNGA sample to the nearest spherical void centre, normalizing the distances by the void radius. Galaxies with a distance $d<r_{\rm void}$ were classified as void galaxies, while those with $d>r_{\rm void}$ were categorized as non-void galaxies. 
Galaxies whose nearest void is close to the catalogue edges were excluded from the analysis, specifically those at a distance of $d<1.5 r_{\rm void}$ from the edge.
This selection leaves us with 176 void galaxies and 3191 non-void galaxies.
We include in Appendix \ref{sec:Samples-characterization} the distributions of redshift, effective radius, and inclination angle of the galaxies in each sample to better characterize them.

\begin{figure}
	\includegraphics[width=\linewidth]{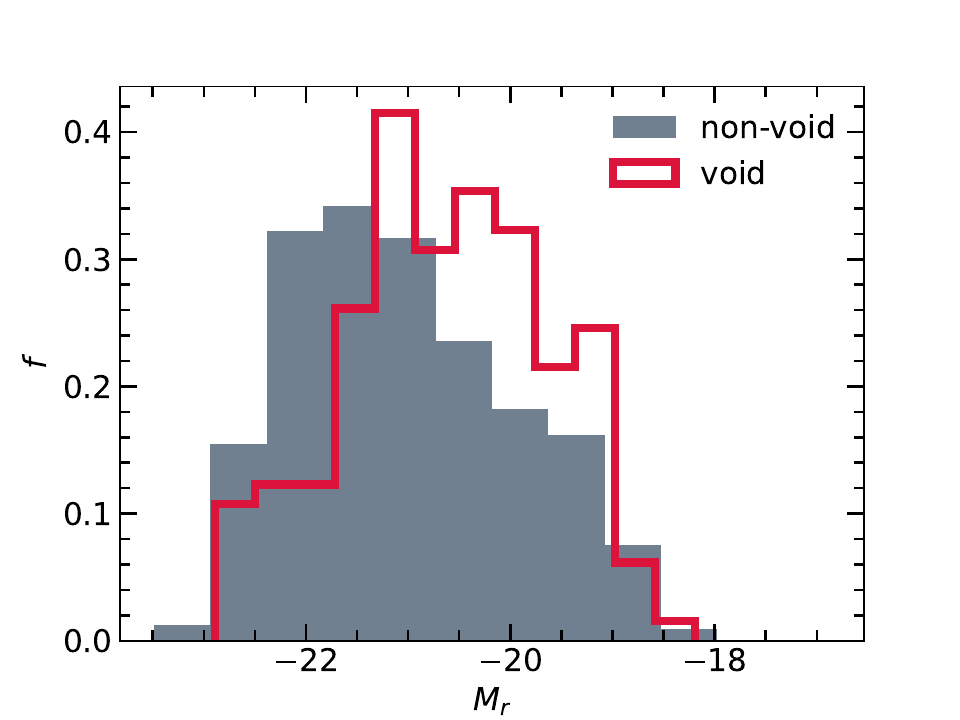}
    \caption{
    Distribution of r-band absolute magnitudes for the galaxy samples. The filled grey histogram represents the non-void galaxy sample, while the red histogram shows the distribution for the void-galaxy sample.
    }
    \label{fig:magnitude_dist}
\end{figure}

\section{Results}
\label{sec:results}

In this section we present the results of our analysis. We focus on the stellar metallicity, age, and gas content of galaxies, while also examining general properties such as colours and star formation rates (SFR). The purpose of focusing on these properties is to compare them with the predictions made in \citet{RodriguezMedrano2024}, where it was found that void galaxies have a higher gas mass, lower stellar metallicity, and younger stellar populations compared to non-void galaxies.

\subsection{General properties}

As mentioned in the introduction, several studies have established that void galaxies exhibit bluer colours and have higher star formation rates (SFR) in comparison with galaxies in denser environments.
Part of this is due to the fact that the magnitude distribution of void galaxies is typically biased toward fainter galaxies, as shown in Figure \ref{fig:magnitude_dist}. The grey histogram corresponds to the r-band absolute magnitude distribution of non-void galaxies, while the red distribution represents the void sample. We can clearly observe the difference in magnitude distribution between the two samples. Since fainter galaxies are typically bluer, star-forming, and younger, it is important, when comparing void and non-void galaxies, to study the galaxies as a function of magnitude or to create matched samples by replicating the magnitude distribution to avoid potential biases. 

We first analyse colours and star formation rates (SFR) of galaxies in our sample, as calculated from MaNGA data

\citep{Sanchez2022}. 

We then extended the analysis to focus on the key properties relevant to this study: stellar metallicity, stellar age, and the gas content of the galaxies.

\begin{figure}
	\includegraphics[width=\linewidth]{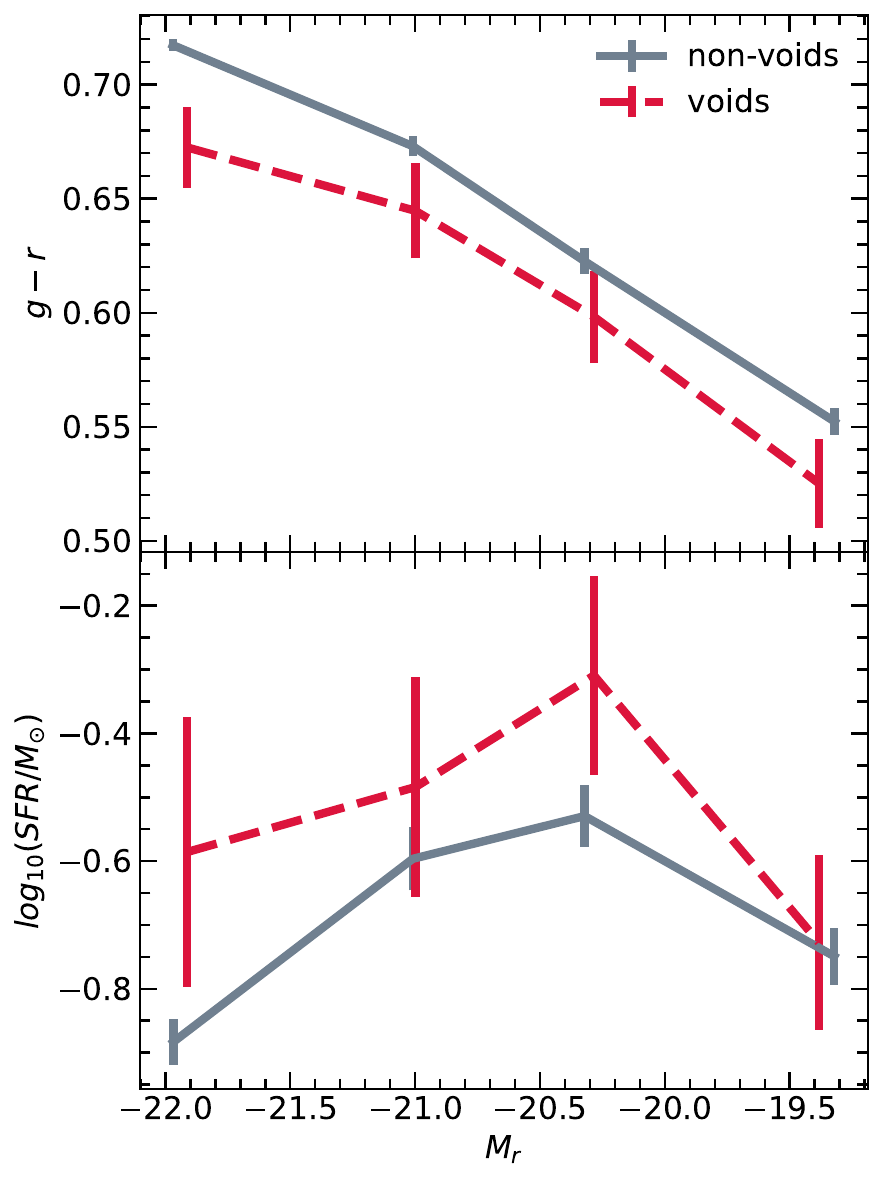}
    \caption{ Astrophysical properties of galaxies as a function of r-band absolute magnitude for void and non-void galaxies. \textit{Top panel:} g-r color; \textit{Bottom panel:} star formation rate (SFR). Both panels show the relation for non-void galaxies (solid grey line) and void galaxies (dashed red line). Error bars indicate the uncertainty associated with the mean value. 
    }
    \label{fig:color-sfr}
\end{figure}

In Fig. \ref{fig:color-sfr}, we present the colour and SFR for our MaNGA dataset. The top panel shows the mean $g-r$ colour as a function of the absolute magnitude in the r-band. The solid grey line indicates the results for the non-void galaxy sample, while the dashed red line represents void galaxies. The shaded regions around each line indicate the uncertainty in the mean, calculated as $s/ \sqrt{n}$ (where $s$ is the standard deviation and $n$ is the sample size). The trend reveals consistently lower $g-r$ values in all magnitude bins for void galaxies, with some differences exceeding the error bars. 
This suggests that void galaxies are systematically bluer. In the bottom panel, we show the SFR.
The SFR is calculated from the luminosity of the $H_{\alpha}$ emission line, which is corrected for extinction caused by interstellar dust and converted into SFR using an empirical relation \citep{Kennicutt1998}.
The figure shows a tendency for galaxies in voids to exhibit higher mean SFR values than galaxies in non-voids.

\begin{figure}
	\includegraphics[width=\linewidth]{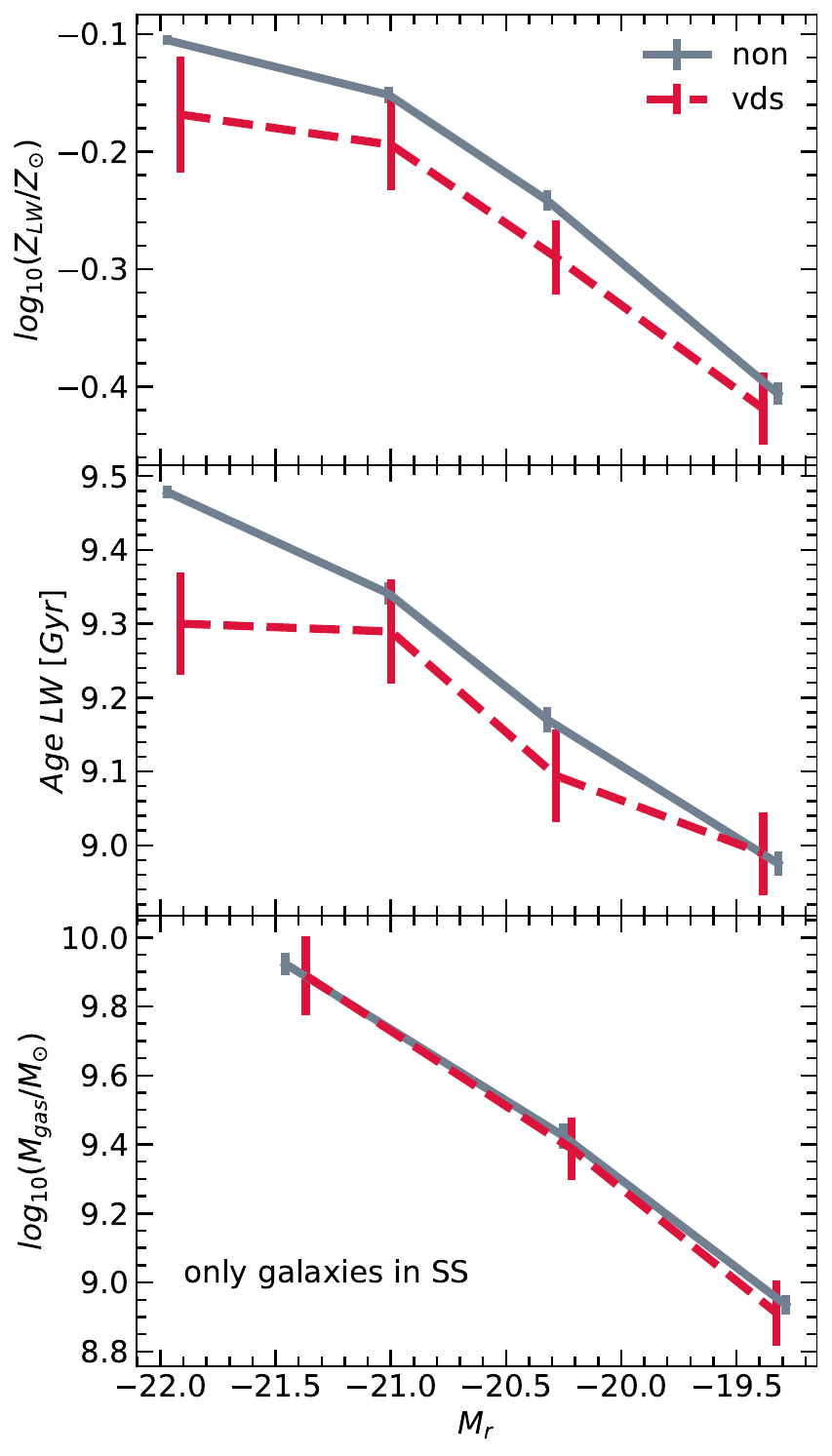}
    \caption{ Astrophysical properties of galaxies as a function of r-band absolute magnitude. \textit{Top panel:} luminosity weighted stellar metallicity; \textit{Middle panel:} luminosity weighted galaxy age; \textit{Bottom panel:} gas mass for galaxies in the secondary sample (SS). The solid grey line represents the relation for the non-void galaxy sample, while the dashed red line represents the relation for the void galaxy sample. Error bars indicate the uncertainty associated with the mean value. 
    }
    \label{fig:met-age-gas}
\end{figure}

Fig. \ref{fig:met-age-gas} shows various properties of the galaxies as a function of absolute r-band magnitude. In all panels, the dashed red line corresponds to void galaxies, and the solid grey line to non-void galaxies. 
For intensive properties such as stellar age and metallicity, we use the value at the effective radius ($R_e$), derived from the azimuthally averaged radial profiles. These profiles are constructed using elliptical annuli out to the maximum radius covered by the FoV \citep{Sanchez2022}. The value at $R_e$ is taken as a representative proxy for the galaxy-wide average, as supported by previous works \citep{Moustakas2010, Sanchez2016aRMxAA}. This approach mitigates biases due to varying spatial coverage among galaxies in the sample.

The top panel displays the luminosity-weighted stellar metallicity at the effective radius $R_e$. 
Across the entire range of magnitudes analysed, void galaxies are systematically less metal-rich than non-void galaxies. These differences have a statistical significance of approximately $\sim 1\,\sigma$, where $\sigma$ is the error in the mean.
The middle panel shows the luminosity-weighted age. For the faintest magnitude bin, there appears to be no statistically significant difference between void and non-void galaxies. However, for the remaining bins, the void galaxy sample consistently shows lower average ages than the non-void sample.
The bottom panel presents the mean gas mass. 
Dust extinction serves as an indicator of the molecular gas content through its connection with the dust-to-gas ratio \citep{Brinchmann2004}.
The gas mass is estimated by integrating the molecular gas surface density ($\Sigma_\text{mol}$) across the field of view (FoV) of each IFU, where $\Sigma_\text{mol}$ is derived from the spaxel-by-spaxel $A_{V,\text{gas}}$ parameter using the linear calibrator from \citep{BarreraBallesteros2021a}.
Given that the gas content is estimated by summing the contributions across the entire FoV, galaxies in the Primary Sample (PS), having a smaller FoV, may be biased toward lower gas content compared to galaxies in the Secondary Sample (SS), which have a larger FoV. For this reason, we decided to present gas estimates only for the Secondary Sample. The presented relation was calculated using only 79 galaxies in voids (1193 for non-void), which led us to divide the sample into three equal-number bins to increase the number of galaxies per bin.
In this case, Fig.\ref{fig:met-age-gas} shows that the estimated mean gas content in galaxies does not differ with environment.

\subsection{Morphology}

It is well known that galaxy morphology is closely related to other galactic properties and that the environment plays a significant role in shaping galaxy morphology. We separate galaxies into early- and late-type categories to analyse the differences between galaxies in void and non-void environments across different morphological types. 

We used the visual classification from \citet{Vazquez-Mata2022}. This classification is based on image mosaics combining r-band images from SDSS and the DESI Legacy Surveys, identifying 13 Hubble types. Each galaxy is assigned a numerical code (T-type) of -5, -2, or a value between 0 and 10, corresponding to its most probable morphological type (E, S0, S0a, Sa, Sab, Sb, Sbc, Sc, Scd, Sd, Sdm, Sm, and Irr). Since we do not have a sufficiently large number of void galaxies to analyze each type individually, we generalized the sample by dividing it into early-type galaxies (ETGs) ($\rm T-type\leq0$) and late-type galaxies (LTGs) ($\rm T-type\geq1$). We do not consider $\rm T-type = 10$, which corresponds to irregular galaxies.

\begin{figure}
   \includegraphics[width=\linewidth]{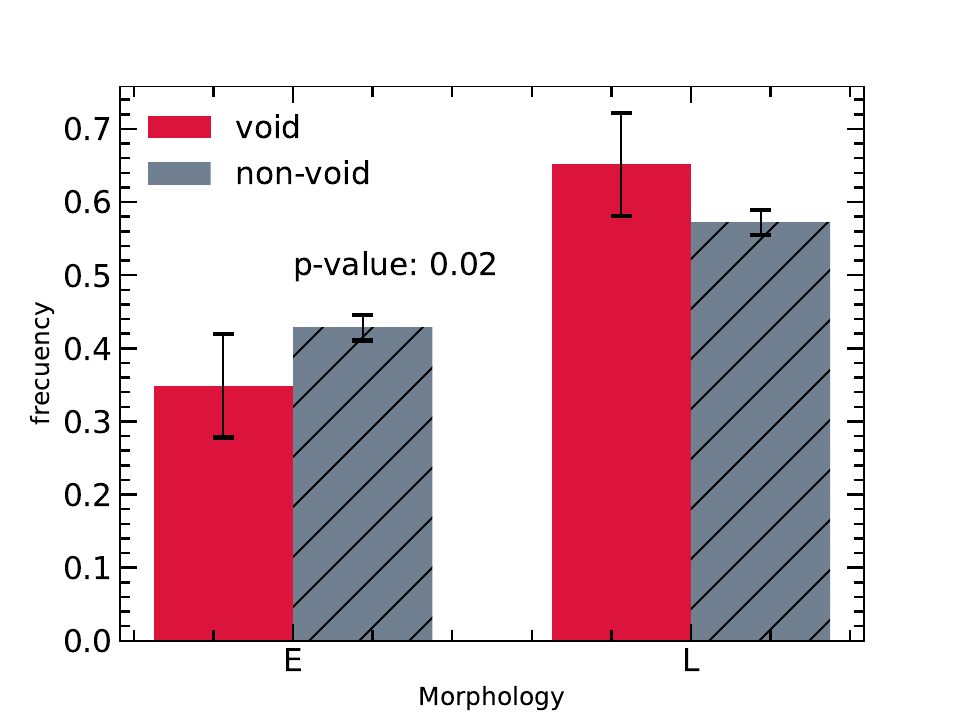}
    \caption{ 
    Fraction of early-type (ETGs) and late-type galaxies (LTGs) in the void (red) and non-void (dashed grey) galaxy samples. Error bars represent the $95\%$ confidence intervals for each proportion. The figure also includes the p-value from a hypothesis test evaluating whether the proportions of early- and late-type galaxies differ significantly between void and non-void environments.
    }
    
    \label{fig:bar_plot-morphology}
\end{figure}

In Fig. \ref{fig:bar_plot-morphology}, we present the results of this classification. The left bars show the proportions of galaxies classified as early types, while the right bars represent late-type galaxies. Void galaxies are shown in red, and non-void galaxies are indicated with dashed grey bars. The error bars represent the $95\%$ confidence intervals for each proportion. 
We find that in our sample of galaxies, $\sim35\%$ ($\sim65\%$) of void galaxies are ETGs (LTGs), while $\sim43\%$ ($\sim57\%$) are ETGs (LTGs) in the non-void sample.
The frequencies for void galaxies suggest a higher fraction of late-type galaxies and a lower fraction of early types. 

To determine if these differences are significant, we tested whether the observed fraction of early-type (or late-type) galaxies in void environments could be due to random variation when compared to non-void environments. The test yields a p-value of $0.02$, supporting the conclusion that both proportions differ significantly.

\begin{figure}
	\includegraphics[width=\linewidth]{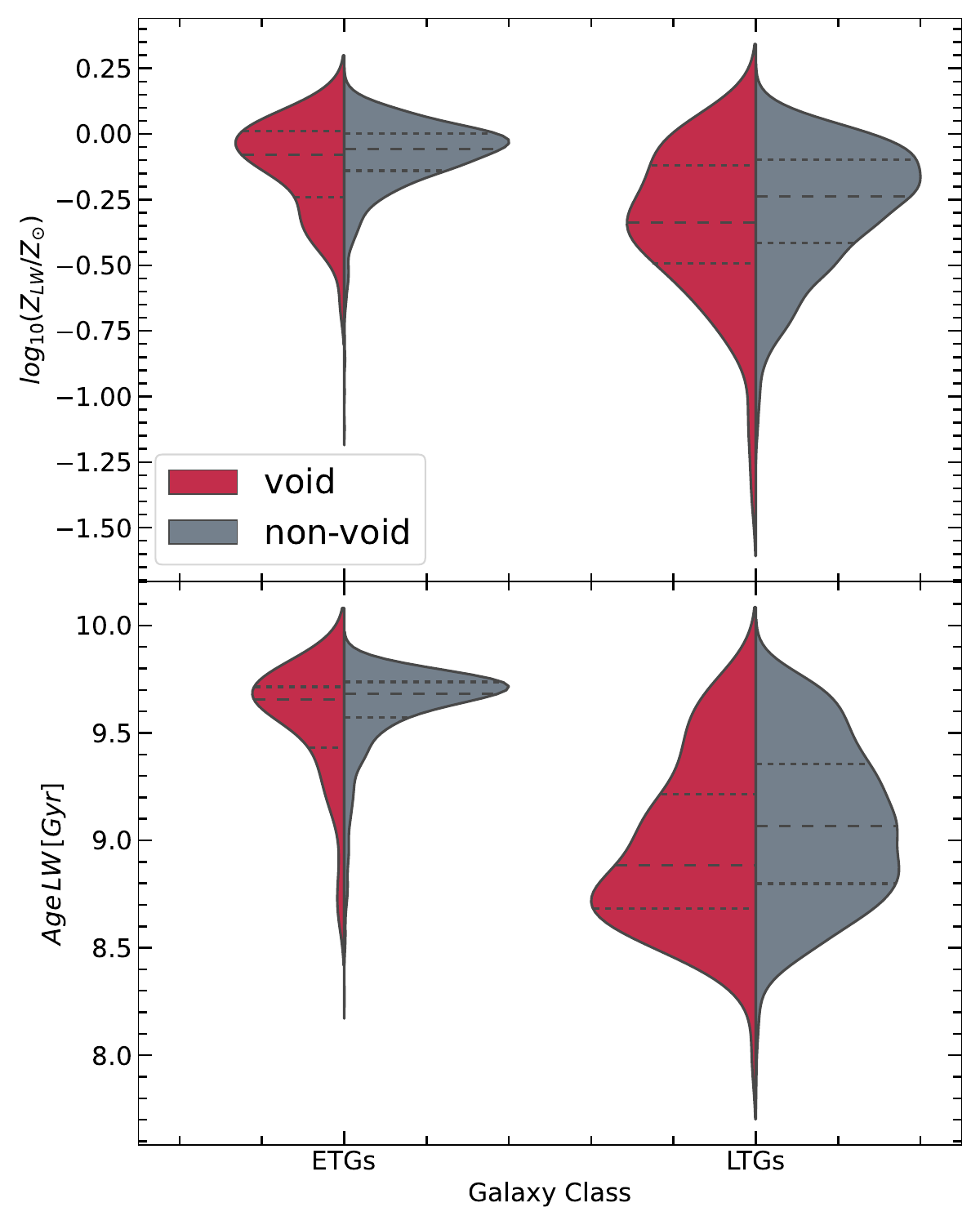}
    \caption{ 
    Distribution of stellar properties for void and non-void galaxies, separated by early-type (ETGs) and late-type (LTGs) galaxies. \textit{Top panel:} stellar metallicity; \textit{Bottom panel:} galaxy age. In each panel, the left side shows the distributions for ETGs, and the right side shows the distributions for LTGs. In each violin plot, the left (red) side represents the void galaxy sample, while the right (grey) side represents the non-void sample. Solid lines within the violin plots indicate the median values and the 25th–75th percentiles.   
    }
    \label{fig:violin_Z-Age}
\end{figure}

After separating the galaxies by morphology in both void and non-void samples, we studied their light-weighted stellar metallicity and age at the effective radius. In Fig. \ref{fig:violin_Z-Age}, we present violin plots for metallicity (top panel) and age (bottom panel). In each panel, the left distributions correspond to ETGs, while the right distributions represent LTGs. The void galaxy sample is shown in red, and the non-void sample in grey. The dashed lines in the violin plots indicate the quartiles. 
The distributions show that ETGs are typically more metal-rich and host older stellar populations compared to LTGs \citep{Kauffman2003, Gallazzi2005}.

For ETGs, although the void and non-void distributions show the same median values for age and metallicity, we observe a tail in the void sample distributions extending towards lower values for both properties.
In the case of LTGs, the distributions appear significantly different. The metallicity distributions suggest a bimodality for the void sample, with a low-metallicity peak being more pronounced. Regarding the age distribution, the void sample is skewed towards younger ages in comparison with the non-void sample.
The medians in both panels indicate that the void sample has lower metallicities and younger ages for LTGs than the non-void galaxy sample.

We remark that aside from morphology, we also investigated whether the gas content of galaxies varies with environment for both morphological types. Our analysis did not reveal any significant differences in gas content between void and non-void galaxies, leading us to focus our discussion and figures on the stellar properties instead. 

\subsection{Age and metallicity distributions}

In the previous subsection, we demonstrated that the distribution of ETGs and LTGs shows differences associated with both morphological type and environment. However, it is important to emphasise that the r-band absolute magnitude distributions associated with the void and non-void samples differs. As shown in Fig. \ref{fig:magnitude_dist}, the void sample distribution is skewed toward fainter magnitudes compared to the non-void sample.
Typically, the brighter a galaxy is, the richer its stellar population is in metals. Then, the observed differences in metallicity and age in Fig. \ref{fig:violin_Z-Age} could be more closely related to differences in the magnitude of galaxies rather than their environmental location. To avoid this bias, we generated a new non-void galaxy sample that matches the r-band absolute magnitude distribution of the void sample. Since the non-void sample contains many more galaxies, we can replicate the magnitude distribution with 10 times more galaxies, which helps to reduce statistical errors.

\begin{figure*}[t]
\includegraphics[width=\textwidth]{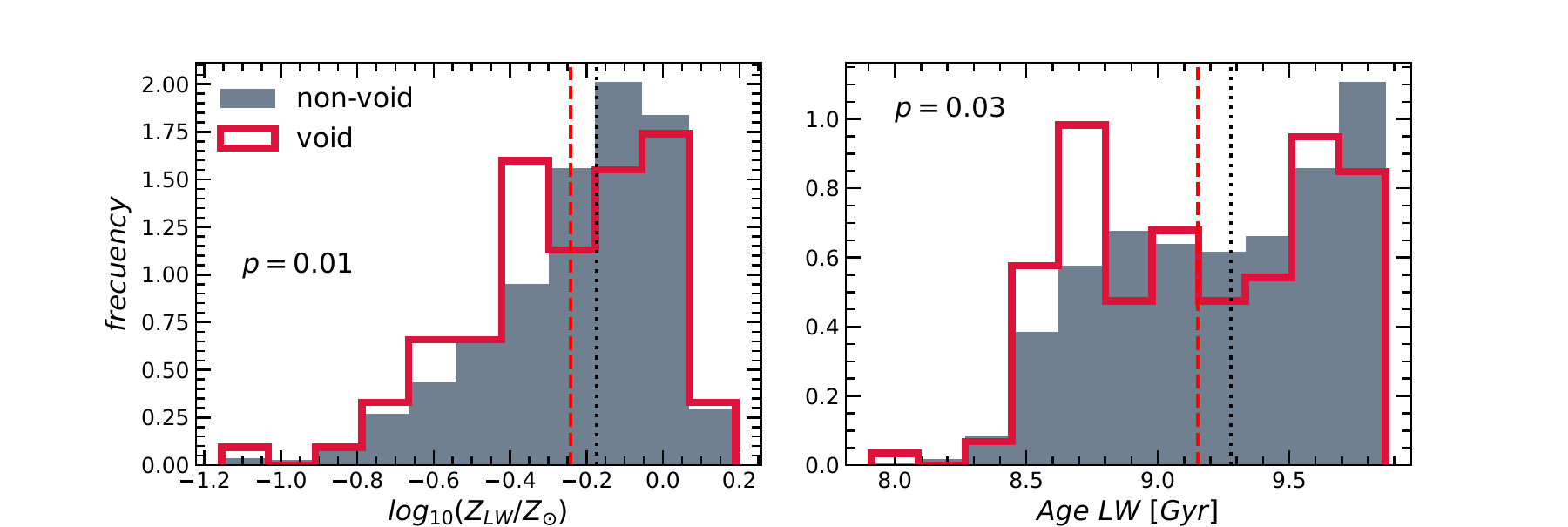}
    \caption{ The left panel shows the distribution of stellar metallicity for a sample of non-void galaxies that mimics the r-band absolute magnitude distribution of void galaxies. The right panel shows the galaxy age distribution for the same samples. The filled grey histogram represents the distribution for non-void galaxies, while the red histogram represents the distribution for void galaxies. A vertical dotted black line indicates the median for the non-void sample, and a dashed red line indicates the median for the void sample in each panel. Each panel also includes the p-value from the Kolmogorov-Smirnov (KS) statistical test. }
    \label{fig:distribuciones_Z-age}
\end{figure*}

In Fig. \ref{fig:distribuciones_Z-age}, we present the metallicity distribution (left panel) and age distribution (right panel) for void galaxies (red histograms) and the new non-void sample (grey histograms) that mimics the r-band magnitude void distribution. Both distributions indicate that void galaxies are less metal-rich and younger than non-void galaxies. This is reflected in the median values shown in the plots and the p-values resulting from a KS test, which indicate that, in both cases, the distributions for void and non-void galaxies are statistically different.

\subsection{Galaxy gradients}

Following the methodology outlined in the previous section, we analyze the metallicity and age profiles of ETGs and LTGs, ensuring that the absolute magnitude distribution of void galaxies matches that of non-void galaxies. The profiles were constructed using the data presented in \citet{Riffel2023} (MEGACUBES) and explained in Section \ref{sec:data_gradients}, where stellar metallicities and ages for each galaxy were calculated in concentric bins centred on the galaxy, extending out to $2\,R_e$ or $1.5\,R_e$ depending on whether the galaxy belongs to the primary or secondary sample \citep{Yan2016, Wake2017}.

We mimic the r-band magnitude distribution of void galaxies using non-void galaxies and stack all galaxies in both samples. The profiles are computed in units of the effective radius, which allows us to normalize galaxy sizes.  
From these stacked galaxies, we calculate the mean value in each bin to construct the mean profiles. Specifically, the mean value and associated error are calculated at radii of $0.5$, $1.0$, $1.5$, and $2.0\, R_e$. Since some galaxies have their properties measured only up to $1.5 R_e$ (mainly those from primary sample), the outermost bin is derived from a reduced subset of galaxies.

\begin{figure*}[h]
    \centering
    \begin{subfigure}{0.45\textwidth}
        \centering
        \includegraphics[width=\textwidth]{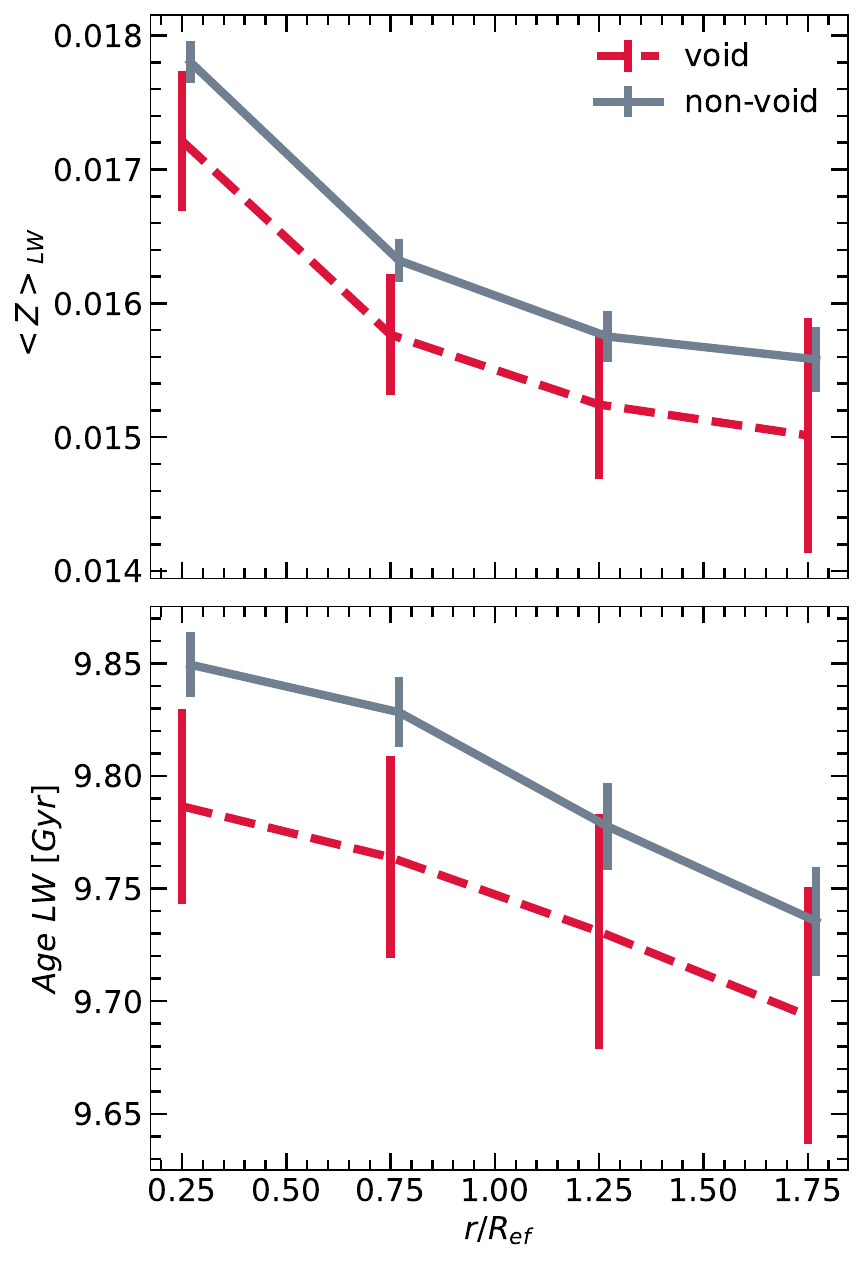}
        \caption{ETGs}
        \label{subfig:gradientes_Z-age_early}
    \end{subfigure}
    \hfill
    \begin{subfigure}{0.45\textwidth}
        \centering
        \includegraphics[width=\textwidth]{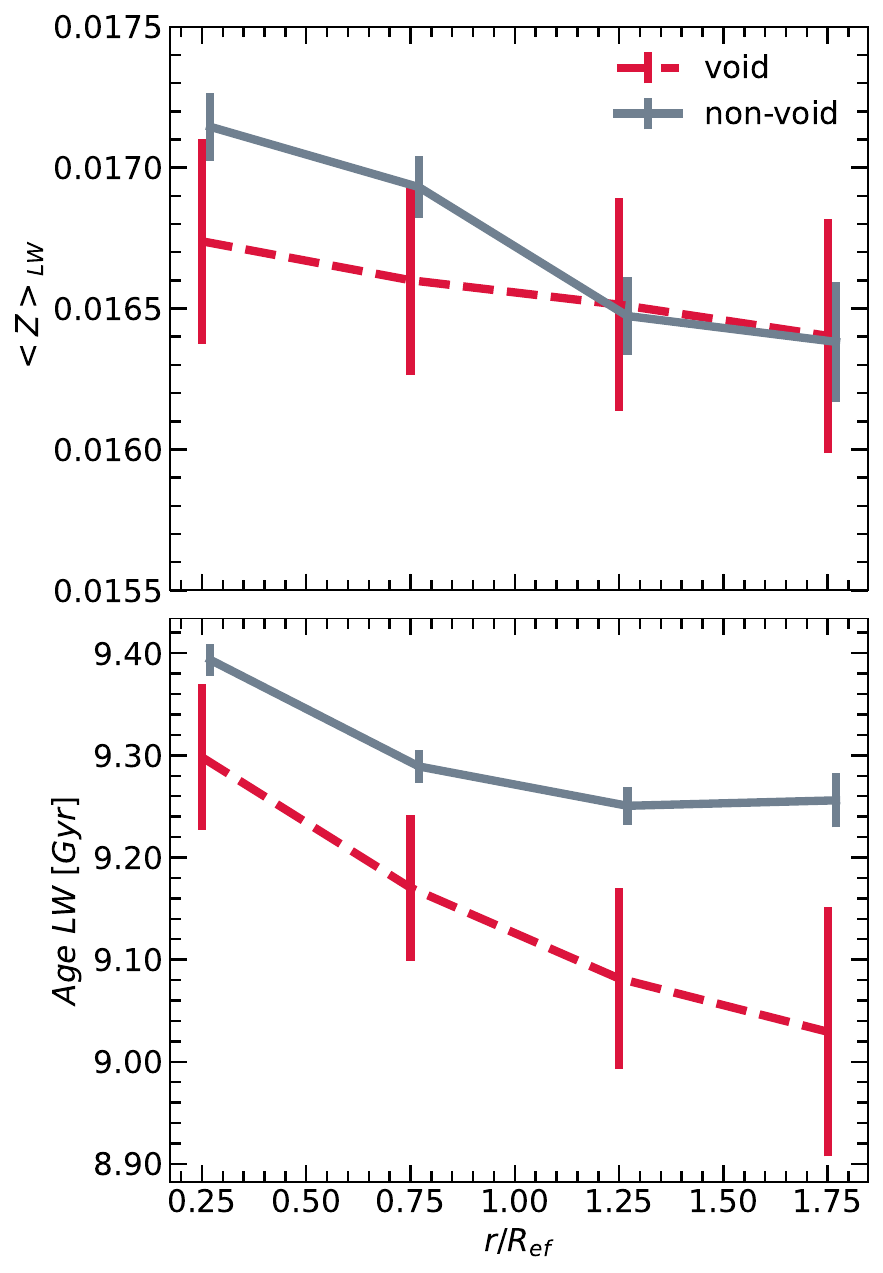}
        \caption{LTGs}
        \label{subfig:subfig:gradientes_Z-age_early}
    \end{subfigure}
    \caption{Profiles of the mean stellar metallicity (top panels) and age (bottom panels). The left columns show the profiles for late-type galaxies (LTGs), and the right columns show the profiles for early-type galaxies (ETGs). Void galaxies are represented in dashed red line, while non-void galaxies are shown in solid grey line. The error associated with the mean is calculated as $\sigma/\sqrt(n)$.  
    }
    \label{fig:gradientes_Z_age}
\end{figure*}

The results of these galaxy profiles are presented in Fig. \ref{fig:gradientes_Z_age}. The left panels show results for ETGs, while the right panels correspond to LTGs.  

For ETGs, in the top panel, we can see that void galaxies systematically exhibit lower metallicity values compared to non-void galaxies. The same is observed for stellar age. On average, for both metallicity and ages, the results suggest a difference between void and non-void galaxies with a significance of approximately $\sim 1\sigma$.
For LTGs, we can see in the top right panel that void galaxies exhibit a slight negative metallicity gradient, while the non-void sample shows a more pronounced negative gradient in the profile, with more metal-rich bulges. However, the differences in the mean profiles lie within the error bars of the void sample. The bottom panel shows the stellar ages, where we observe a negative gradient for both void and non-void galaxies. The figure indicates that void galaxies systematically have lower stellar ages, particularly significant in the outer regions of the disk, where the differences exceed the associated error.
We observe, regardless of the environment, that both early and late-type galaxies exhibit negative age and metallicity gradients, consistent with the inside-out scenario \citep{PerezE2013, GonzalezDelgado2014, GarciaBenito2017}.

For a reference comparable with other studies, we computed the metallicity gradients $\nabla [Z/H]$ and age gradients for our samples using the 0.25-1.25 $R_e$ interval. In the case of metallicity, the gradients were calculated on a logarithmic scale, adopting a solar metallicity of $Z_\odot = 0.017$. For LTGs, the metallicity gradients are $\nabla [Z/H]= 0.006 \pm 0.014$ (void sample) and $\nabla [Z/H]= -0.016 \pm 0.005$ (non-void sample). For ETGs, the metallicity gradients are $\nabla [Z/H]= -0.053 \pm 0.021$ (void sample) and $\nabla [Z/H]= -0.053 \pm 0.006$ (non-void sample). All these metallicity gradients have units of $[\rm dex/R_e]$.  
For the age gradients, we obtained the following values for LTGs: $\nabla \text{Age} = -0.217 \pm 0.005$ (void sample) and $\nabla \text{Age} = -0.142 \pm 0.001$ (non-void sample). For ETGs: $\nabla \text{Age} = -0.056 \pm 0.003$ (void sample) and $\nabla \text{Age} = -0.063 \pm 0.001$ (non-void sample). In this case, the units of these gradients are $[\rm log_{10}(Gyr)/R_e]$.

For each of MaNGA megacubes, in \citet{Riffel2023} the authors calculated the \textit{binned population vectors}, 
which represent the proportion of the stellar population in different age ranges (percentage of stellar population). These are defined as follows:

\begin{itemize}
    \item xyy light:  $t \leq 10\, Myr$
    \item xyo light: $14 \,Myr <t \leq 56\, Myr$
    \item xiy light: $100\, Myr < t \leq 500\, Myr$
    \item xii light: $630\, Myr < t \leq 800\, Myr$
    \item xio light: $890\, Myr < t \leq 2.0\, Gyr$
    \item xo light: $5.0\, Gyr < t \leq 13\, Gyr$
    
\end{itemize}

The letters represent: y (young), i (intermediate), and o (old). To simplify our results, we group the vectors into three categories: young, intermediate, and old, where: 

\begin{itemize}
    \item young: $t \leq 56\, Myr$
    \item intermediate: $100\, Myr < t \leq 2\,Gyr$
    \item old: $5\, Gyr < t \leq 13\, Gyr$
\end{itemize}

\begin{figure*}[h]
    \centering
    \begin{subfigure}{0.45\textwidth}
        \centering
        \includegraphics[width=\textwidth]{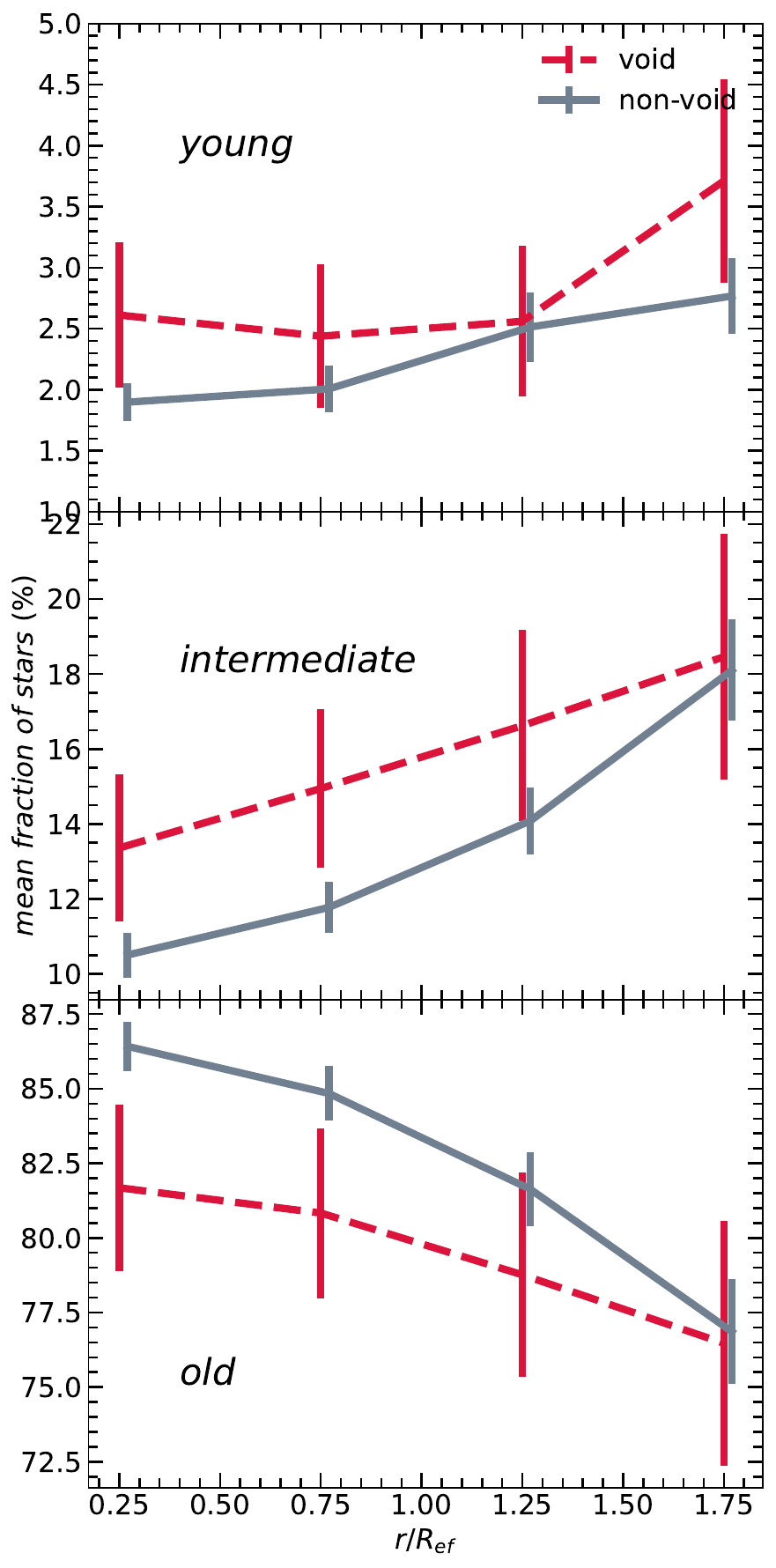}
        \caption{ETGs}
        \label{subfig:early-type (ETGs)}
    \end{subfigure}
    \hfill
    \begin{subfigure}{0.45\textwidth}
        \centering
        \includegraphics[width=\textwidth]{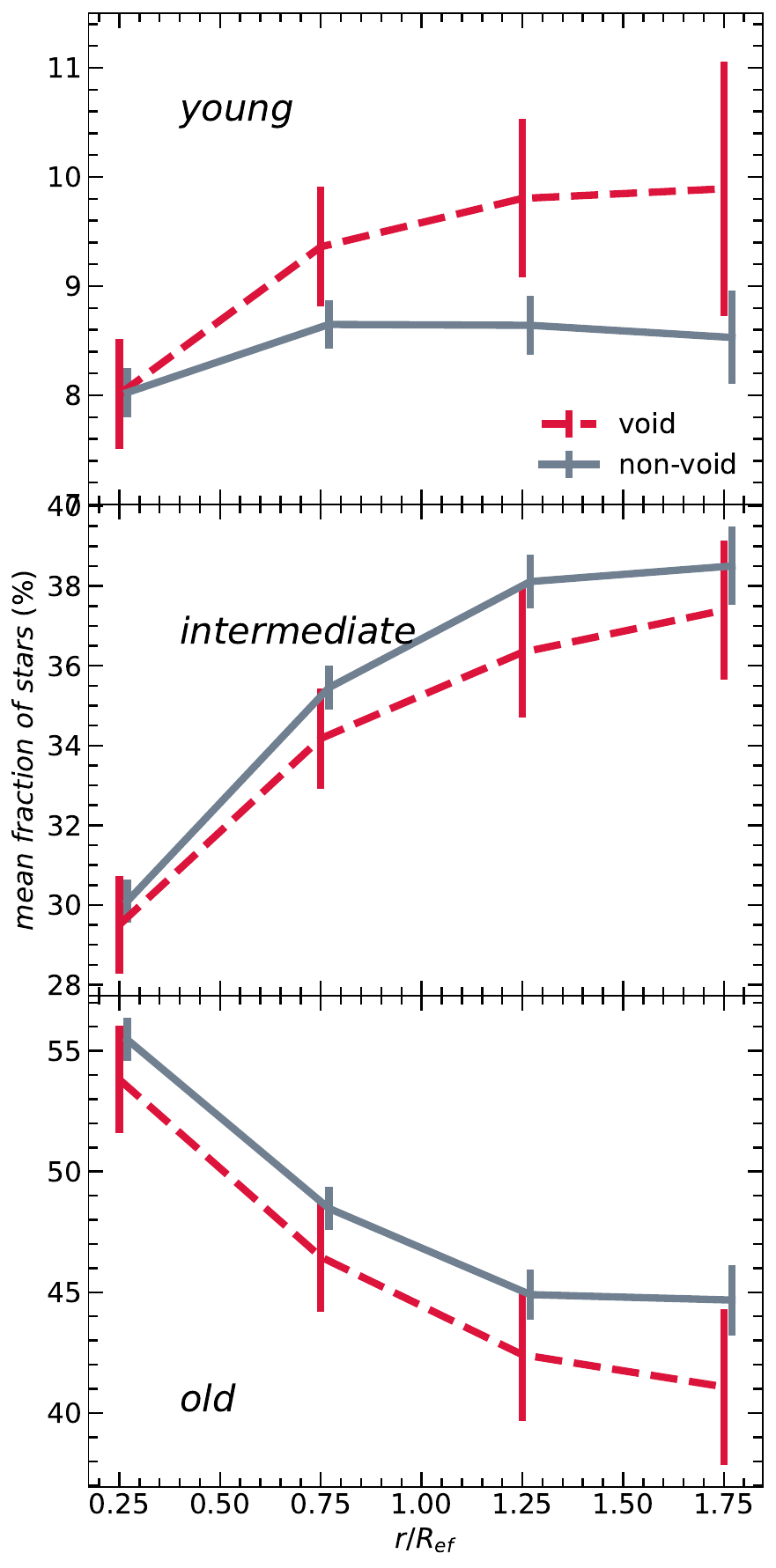}
        \caption{LTGs}
        \label{subfig:late-type (LTGs) }
    \end{subfigure}
    \caption{The figure shows the profiles of mean fraction of stars in different age bins for early-type (left column) and late-type galaxies (right column). The top panel shows the young population (\(t < 56\) Myr), the central panels show the intermediate population (\(100 \, \text{Myr} < t < 2 \, \text{Gyr}\)), and the bottom panels show the old population (\(2 \, \text{Gyr} < t < 5 \, \text{Gyr}\)). Void galaxies are represented in dashed red, and non-void galaxies are shown in grey.
    The error associated with the mean is calculated as $\sigma/\sqrt(n)$.}
    \label{fig:gradientes_populations}
\end{figure*}

In Fig. \ref{fig:gradientes_populations}, we show the population profiles for these three categories, again separated by morphological type (ETGs in the left panels, LTGs in the right). 
The top panels show the young population. 
For ETGs, the figure shows a younger population in all bins for void galaxies. The intermediate-age population is also higher in the void sample compared to the non-void sample. The opposite occurs for the old stellar population, where we observe higher fractions in the non-void sample. The trends are consistently observed, indicating younger stellar populations in void galaxies across all radial bins.

For LTGs, the figure shows that the the profile for void galaxies exhibits a higher fraction of young populations compared to the non-void sample. For the intermediate (central panel) population, we do not find differences between the void and non-void samples. In the old populations (bottom panel), we observe a trend indicating that the non-void sample displays higher fractions of these populations than the void sample.

\section{Discusions}
\label{sec:discusions}

Our results with MaNGA data are in line with previous studies analyzing the characteristics of galaxies in cosmic voids from different surveys. However, our findings provide a new perspective by utilizing integral field spectroscopy (IFS) data from the MaNGA survey, allowing for more detailed modelling of stellar populations compared to traditional fibre spectroscopy or photometrical methods.
In our analysis, we observed that void galaxies have lower absolute magnitudes and are bluer than galaxies in denser environments. We also identified a trend indicating that void galaxies exhibit higher star formation rates. All these results are in agreement with the general consensus of the void galaxy properties \citep{Rojas2004, Patiri2006b, Tavasoli2015, RodriguezMedrano2023}.

Our findings also show that, as a function of r-band absolute magnitude, void galaxies tend to have lower stellar metallicities than non-void galaxies. For the brigther galaxies, the difference is not statistically significant, which indicate that the metallicities are similar in both environments. In \citet{RodriguezMedrano2024}, we analysed the void galaxy population in the IllustrisTNG simulation and found a similar result: void galaxies exhibit lower stellar metallicities, with the signal decreasing as stellar mass increases. 
The same signal was found in SDSS-DR7 data in \citep{Dominguez2023a}.
Given that stellar mass correlates with absolute magnitude, our results in this paper are consistent with those in previous works.

We also explored whether void galaxies have higher gas content compared to galaxies in denser environments, as suggested in \citet{RodriguezMedrano2024}. However, we did not find a dependence of gas content on environment. Across the entire range of absolute magnitudes analysed, galaxies displayed same mean gas masses. 
One of the main results from the work of \citep{Florez2021} is that void galaxies are richer in gas, even when controlling for the morphology and mass of the galaxies. In this study, the gas mass is derived either directly from $H_I$ data for some galaxies or indirectly through a relation between the galaxy color and the gas mass fraction. Beyond the differences in the determination of gas mass between both studies, the definition of voids also differs. In their work, galaxies are classified as void and non-void based on their distance to the nearest neighboring galaxies, making this classification dependent on the "local environment," in contrast to our classification, which is based on the integrated large-scale density field. 
Another study investigating the gas content in void galaxies is \citep{DominguezGomez2022}, where no significant differences in gas mass are found between a sample of void galaxies and a control sample of galaxies in denser environments. In this study, the void galaxy sample is selected based on the large-scale density field, making their void galaxy definition more comparable to that of our samples.  
Despite the differences among the various studies in the literature, it would be interesting to have a sample of fainter galaxies to carry out these analyses, as the differences in gas content predicted in our numerical work \citet{RodriguezMedrano2024} are expected to be more significant in low-mass galaxies, according to our previous studies.
On the other hand, we emphasize that the gas mass estimates used in this work depend on properties such as galaxy inclination and star formation rate (SFR) \citep{BarreraBallesteros2021a}. A detailed study of the gas content in galaxies should not only control for global magnitudes, as we have done here, but also ensure that the samples are unbiased with respect to inclination and SFR. Considering these aspects, and the fact that we used only the secondary sample to ensure consistent spatial coverage across galaxies, we believe that the results related to gas mass should be interpreted with appropriate caution.

As the introduction states, a high proportion of late-type galaxies (LTGs) is expected in cosmic voids \citep{Hoyle2005, Ricciardelli2017}. In our data, we found that the fraction of LTGs is slightly higher in void environments ($65 \%$ and $57\%$).
We obtain a p-value $p=0.02$ which support the hypothesis of a larger fraction of LTGs in voids in comparison with denser environments.

When we analyse the distribution of metallicity by morphological type (Fig. \ref{fig:violin_Z-Age}), we found that early-type galaxies (ETGs) in both void and non-void environments generally exhibit high light-weighted stellar metallicities than LTGs.

However, the metallicity distribution for void environments displays a bimodal behavior, with a smaller population of galaxies at lower metallicities. In the case of LTGs, void galaxies tend to have lower stellar metallicities, also exhibiting a bimodal distribution.
For the light-weighted age (LW-age) distributions, ETGs showed similar patterns in both environments. In contrast, for LTGs, the age distributions differed between environments, with void galaxies skewed towards younger ages, indicating a younger population of galaxies.

These differences in age and metallicity in late-type galaxies persist even in samples with the same magnitude distribution, confirming an environmental effect and not simply a bias due to the magnitude of galaxies.

The galaxy metallicity profiles presented in Fig. \ref{fig:gradientes_Z_age} exhibit 
negative gradients for LTGs in both void and non-void samples, with the gradient being more pronounced in non-void galaxies. However, these two behaviours are not statistically distinguishable when considering the errors in the calculated gradients.
For ETGs, we found a trend suggesting a less metal-rich stellar population across the entire range of distances for the void galaxy population.

The age profiles for LTGs and ETGs consistently indicate younger populations in void galaxies. In the case of ETGs, the profiles are similar but display an offset to lower values in the void environment. For LTGs, the gradients appear to differ in the outer regions (discs), indicating similar bulges in both void and non-void environments, but suggesting that the discs in void galaxies host a younger population.

A recent study on the relationship between age gradients and the environment of voids is presented by \citet{Conrado2024}. In that work, the authors use a sample of void galaxies and a control sample, classifying them morphologically as E, S0, Sa, Sb, Sc, and Sd. When comparing both samples, they find that for all morphological types, the age profiles indicate that void galaxies have younger stellar populations. In our case, due to the low number of galaxies, we decided to classify them into only two morphological types (ETGs and LTGs). Nevertheless, we found a qualitative agreement between our general results and those obtained by \citet{Conrado2024}.

Overall, our results indicate that void galaxies are younger and possess lower stellar metallicities. The age findings are consistent with those obtained from simulations \citep{Tonnensen2015, Alfaro2020, RodriguezMedrano2022, RodriguezMedrano2024} and observational data \citep{Dominguez2023b, Torres-Rios2024}, where the authors employed spectral distribution models on spectroscopic data. This observed difference in galaxy ages suggests that the lower metallicities in void galaxies may be attributed to their younger stellar populations.

Finally, we want to acknowledge some limitations of our study.
The MaNGA galaxy sample, while the most extensive survey conducted to date using integral field spectroscopy (IFS), is limited by its bias toward brighter galaxies, which may affect our results regarding gas content and the properties of lower-mass galaxies. Furthermore, although the morphological separation into early and late types provides clarity, morphology can be classified in greater detail based on the Hubble sequence or different T-types. However, we emphasize that our decision to focus solely on two morphological types was driven by the need to achieve a statistically significant number of galaxies within each type, thereby enhancing the robustness of our results.

\section{Conclusions}
\label{sec:conclusions}
In this paper, we analyse IFS data from the MaNGA survey, separating galaxies into those located within cosmic voids and those outside. The main findings of this work can be summarised as follows: 
\begin{itemize}
    \item At a given absolute magnitude, void galaxies are younger and exhibit smaller metallities than those in non-void environement.
    
    \item The gas content of galaxies in voids and non-voids environments is similar.
    \item When matching the absolute magnitude distribution of void galaxies with that of non-void galaxies, we find statistically significant evidence that the stellar metallicity and age distributions are different between the two samples. The comparison shows that void galaxies are younger and less metal-rich.
    \item These differences in evolutionary signals are evident for both early- and late-type galaxies.
\end{itemize}

The results related to the metallicity and age of galaxies confirm some of the key findings obtained through numerical simulations in \citet{RodriguezMedrano2024}. However, we did not find differences in the gas content of void galaxies, which was one of the objectives of this study. This could be due to the need for a sample of lower-mass galaxies than those used in this work.

In conclusion, we show in this work how the void environment affects galaxy evolution, influencing star formation activity. Our results suggest a delayed evolutionary path for galaxies in large-scale underdense environments, which is reflected in a younger stellar population and lower metallicities.

\begin{acknowledgements}

    The authors thank the referee for reviewing the manuscript and for their helpful comments and suggestions. ARM thanks Ornela Marioni for the valuable discussions held during the preparation of this work. 
      This work was partially supported by grants PICT 2021-00442 awarded by Fondo para la Investigación Científica y Tecnológica (FONCYT), grant PIP 2022-11220210100520 by the Consejo de Investigaciones Científicas y Técnicas de la República Argentina (CONICET), and the Secretaría de Ciencia y Técnica de la Universidad Nacional de Córdoba (SeCyT). ARM is doctoral fellow of CONICET. DJP, DM, ANR, and FAS are members of the Carrera del Investigador Científico (CONICET).

 This project makes use of the MaNGA-Pipe3D dataproducts. We thank the IA-UNAM MaNGA team for creating this catalogue, and the Conacyt Project CB-285080 for supporting them.
      
Funding for the Sloan Digital Sky 
Survey IV has been provided by the 
Alfred P. Sloan Foundation, the U.S. 
Department of Energy Office of 
Science, and the Participating 
Institutions. 

SDSS-IV acknowledges support and 
resources from the Center for High 
Performance Computing  at the 
University of Utah. The SDSS 
website is www.sdss4.org.

SDSS-IV is managed by the 
Astrophysical Research Consortium 
for the Participating Institutions 
of the SDSS Collaboration including 
the Brazilian Participation Group, 
the Carnegie Institution for Science, 
Carnegie Mellon University, Center for 
Astrophysics | Harvard \& 
Smithsonian, the Chilean Participation 
Group, the French Participation Group, 
Instituto de Astrof\'isica de 
Canarias, The Johns Hopkins 
University, Kavli Institute for the 
Physics and Mathematics of the 
Universe (IPMU) / University of 
Tokyo, the Korean Participation Group, 
Lawrence Berkeley National Laboratory, 
Leibniz Institut f\"ur Astrophysik 
Potsdam (AIP),  Max-Planck-Institut 
f\"ur Astronomie (MPIA Heidelberg), 
Max-Planck-Institut f\"ur 
Astrophysik (MPA Garching), 
Max-Planck-Institut f\"ur 
Extraterrestrische Physik (MPE), 
National Astronomical Observatories of 
China, New Mexico State University, 
New York University, University of 
Notre Dame, Observat\'ario 
Nacional / MCTI, The Ohio State 
University, Pennsylvania State 
University, Shanghai 
Astronomical Observatory, United 
Kingdom Participation Group, 
Universidad Nacional Aut\'onoma 
de M\'exico, University of Arizona, 
University of Colorado Boulder, 
University of Oxford, University of 
Portsmouth, University of Utah, 
University of Virginia, University 
of Washington, University of 
Wisconsin, Vanderbilt University, 
and Yale University.

\end{acknowledgements}

\bibliographystyle{aa}
\bibliography{bibliografia} 

\begin{thebibliography}{74}
\expandafter\ifx\csname natexlab\endcsname\relax\def\natexlab#1{#1}\fi

\bibitem[{{Abazajian} {et~al.}(2009){Abazajian}, {Adelman-McCarthy},
  {Ag{\"u}eros}, {Allam}, {Allende Prieto}, {An}, {Anderson}, {Anderson},
  {Annis}, {Bahcall}, {Bailer-Jones}, {Barentine}, {Bassett}, {Becker},
  {Beers}, {Bell}, {Belokurov}, {Berlind}, {Berman}, {Bernardi}, {Bickerton},
  {Bizyaev}, {Blakeslee}, {Blanton}, {Bochanski}, {Boroski}, {Brewington},
  {Brinchmann}, {Brinkmann}, {Brunner}, {Budav{\'a}ri}, {Carey}, {Carliles},
  {Carr}, {Castander}, {Cinabro}, {Connolly}, {Csabai}, {Cunha}, {Czarapata},
  {Davenport}, {de Haas}, {Dilday}, {Doi}, {Eisenstein}, {Evans}, {Evans},
  {Fan}, {Friedman}, {Frieman}, {Fukugita}, {G{\"a}nsicke}, {Gates},
  {Gillespie}, {Gilmore}, {Gonzalez}, {Gonzalez}, {Grebel}, {Gunn},
  {Gy{\"o}ry}, {Hall}, {Harding}, {Harris}, {Harvanek}, {Hawley}, {Hayes},
  {Heckman}, {Hendry}, {Hennessy}, {Hindsley}, {Hoblitt}, {Hogan}, {Hogg},
  {Holtzman}, {Hyde}, {Ichikawa}, {Ichikawa}, {Im}, {Ivezi{\'c}}, {Jester},
  {Jiang}, {Johnson}, {Jorgensen}, {Juri{\'c}}, {Kent}, {Kessler}, {Kleinman},
  {Knapp}, {Konishi}, {Kron}, {Krzesinski}, {Kuropatkin}, {Lampeitl},
  {Lebedeva}, {Lee}, {Lee}, {French Leger}, {L{\'e}pine}, {Li}, {Lima}, {Lin},
  {Long}, {Loomis}, {Loveday}, {Lupton}, {Magnier}, {Malanushenko},
  {Malanushenko}, {Mandelbaum}, {Margon}, {Marriner}, {Mart{\'\i}nez-Delgado},
  {Matsubara}, {McGehee}, {McKay}, {Meiksin}, {Morrison}, {Mullally}, {Munn},
  {Murphy}, {Nash}, {Nebot}, {Neilsen}, {Newberg}, {Newman}, {Nichol},
  {Nicinski}, {Nieto-Santisteban}, {Nitta}, {Okamura}, {Oravetz}, {Ostriker},
  {Owen}, {Padmanabhan}, {Pan}, {Park}, {Pauls}, {Peoples}, {Percival}, {Pier},
  {Pope}, {Pourbaix}, {Price}, {Purger}, {Quinn}, {Raddick}, {Re Fiorentin},
  {Richards}, {Richmond}, {Riess}, {Rix}, {Rockosi}, {Sako}, {Schlegel},
  {Schneider}, {Scholz}, {Schreiber}, {Schwope}, {Seljak}, {Sesar}, {Sheldon},
  {Shimasaku}, {Sibley}, {Simmons}, {Sivarani}, {Allyn Smith}, {Smith},
  {Smol{\v{c}}i{\'c}}, {Snedden}, {Stebbins}, {Steinmetz}, {Stoughton},
  {Strauss}, {SubbaRao}, {Suto}, {Szalay}, {Szapudi}, {Szkody}, {Tanaka},
  {Tegmark}, {Teodoro}, {Thakar}, {Tremonti}, {Tucker}, {Uomoto}, {Vanden
  Berk}, {Vandenberg}, {Vidrih}, {Vogeley}, {Voges}, {Vogt}, {Wadadekar},
  {Watters}, {Weinberg}, {West}, {White}, {Wilhite}, {Wonders}, {Yanny},
  {Yocum}, {York}, {Zehavi}, {Zibetti}, \& {Zucker}}]{Abazajian2009}
{Abazajian}, K.~N., {Adelman-McCarthy}, J.~K., {Ag{\"u}eros}, M.~A., {et~al.}
  2009, \apjs, 182, 543

\bibitem[{{Alam} {et~al.}(2015){Alam}, {Albareti}, {Allende Prieto}, {Anders},
  {Anderson}, {Anderton}, {Andrews}, {Armengaud}, {Aubourg}, {Bailey}, {Basu},
  {Bautista}, {Beaton}, {Beers}, {Bender}, {Berlind}, {Beutler}, {Bhardwaj},
  {Bird}, {Bizyaev}, {Blake}, {Blanton}, {Blomqvist}, {Bochanski}, {Bolton},
  {Bovy}, {Shelden Bradley}, {Brandt}, {Brauer}, {Brinkmann}, {Brown},
  {Brownstein}, {Burden}, {Burtin}, {Busca}, {Cai}, {Capozzi}, {Carnero
  Rosell}, {Carr}, {Carrera}, {Chambers}, {Chaplin}, {Chen}, {Chiappini},
  {Chojnowski}, {Chuang}, {Clerc}, {Comparat}, {Covey}, {Croft}, {Cuesta},
  {Cunha}, {da Costa}, {Da Rio}, {Davenport}, {Dawson}, {De Lee}, {Delubac},
  {Deshpande}, {Dhital}, {Dutra-Ferreira}, {Dwelly}, {Ealet}, {Ebelke},
  {Edmondson}, {Eisenstein}, {Ellsworth}, {Elsworth}, {Epstein}, {Eracleous},
  {Escoffier}, {Esposito}, {Evans}, {Fan}, {Fern{\'a}ndez-Alvar}, {Feuillet},
  {Filiz Ak}, {Finley}, {Finoguenov}, {Flaherty}, {Fleming}, {Font-Ribera},
  {Foster}, {Frinchaboy}, {Galbraith-Frew}, {Garc{\'\i}a},
  {Garc{\'\i}a-Hern{\'a}ndez}, {Garc{\'\i}a P{\'e}rez}, {Gaulme}, {Ge},
  {G{\'e}nova-Santos}, {Georgakakis}, {Ghezzi}, {Gillespie}, {Girardi},
  {Goddard}, {Gontcho}, {Gonz{\'a}lez Hern{\'a}ndez}, {Grebel}, {Green},
  {Grieb}, {Grieves}, {Gunn}, {Guo}, {Harding}, {Hasselquist}, {Hawley},
  {Hayden}, {Hearty}, {Hekker}, {Ho}, {Hogg}, {Holley-Bockelmann}, {Holtzman},
  {Honscheid}, {Huber}, {Huehnerhoff}, {Ivans}, {Jiang}, {Johnson},
  {Kinemuchi}, {Kirkby}, {Kitaura}, {Klaene}, {Knapp}, {Kneib}, {Koenig},
  {Lam}, {Lan}, {Lang}, {Laurent}, {Le Goff}, {Leauthaud}, {Lee}, {Lee},
  {Licquia}, {Liu}, {Long}, {L{\'o}pez-Corredoira}, {Lorenzo-Oliveira},
  {Lucatello}, {Lundgren}, {Lupton}, {Mack}, {Mahadevan}, {Maia}, {Majewski},
  {Malanushenko}, {Malanushenko}, {Manchado}, {Manera}, {Mao}, {Maraston},
  {Marchwinski}, {Margala}, {Martell}, {Martig}, {Masters}, {Mathur},
  {McBride}, {McGehee}, {McGreer}, {McMahon}, {M{\'e}nard}, {Menzel},
  {Merloni}, {M{\'e}sz{\'a}ros}, {Miller}, {Miralda-Escud{\'e}}, {Miyatake},
  {Montero-Dorta}, {More}, {Morganson}, {Morice-Atkinson}, {Morrison},
  {Mosser}, {Muna}, {Myers}, {Nandra}, {Newman}, {Neyrinck}, {Nguyen},
  {Nichol}, {Nidever}, {Noterdaeme}, {Nuza}, {O'Connell}, {O'Connell},
  {O'Connell}, {Ogando}, {Olmstead}, {Oravetz}, {Oravetz}, {Osumi}, {Owen},
  {Padgett}, {Padmanabhan}, {Paegert}, {Palanque-Delabrouille}, {Pan},
  {Parejko}, {P{\^a}ris}, {Park}, {Pattarakijwanich}, {Pellejero-Ibanez},
  {Pepper}, {Percival}, {P{\'e}rez-Fournon}, {P{\'e}rez-R{\`a}fols},
  {Petitjean}, {Pieri}, {Pinsonneault}, {Porto de Mello}, {Prada}, {Prakash},
  {Price-Whelan}, {Protopapas}, {Raddick}, {Rahman}, {Reid}, {Rich}, {Rix},
  {Robin}, {Rockosi}, {Rodrigues}, {Rodr{\'\i}guez-Torres}, {Roe}, {Ross},
  {Ross}, {Rossi}, {Ruan}, {Rubi{\~n}o-Mart{\'\i}n}, {Rykoff},
  {Salazar-Albornoz}, {Salvato}, {Samushia}, {S{\'a}nchez}, {Santiago},
  {Sayres}, {Schiavon}, {Schlegel}, {Schmidt}, {Schneider}, {Schultheis},
  {Schwope}, {Sc{\'o}ccola}, {Scott}, {Sellgren}, {Seo}, {Serenelli}, {Shane},
  {Shen}, {Shetrone}, {Shu}, {Silva Aguirre}, {Sivarani}, {Skrutskie},
  {Slosar}, {Smith}, {Sobreira}, {Souto}, {Stassun}, {Steinmetz}, {Stello},
  {Strauss}, {Streblyanska}, {Suzuki}, {Swanson}, {Tan}, {Tayar}, {Terrien},
  {Thakar}, {Thomas}, {Thomas}, {Thompson}, {Tinker}, {Tojeiro}, {Troup},
  {Vargas-Maga{\~n}a}, {Vazquez}, {Verde}, {Viel}, {Vogt}, {Wake}, {Wang},
  {Weaver}, {Weinberg}, {Weiner}, {White}, {Wilson}, {Wisniewski},
  {Wood-Vasey}, {Ye`che}, {York}, {Zakamska}, {Zamora}, {Zasowski}, {Zehavi},
  {Zhao}, {Zheng}, {Zhou}, {Zhou}, {Zou}, \& {Zhu}}]{Alam2015}
{Alam}, S., {Albareti}, F.~D., {Allende Prieto}, C., {et~al.} 2015, \apjs, 219,
  12

\bibitem[{{Alfaro} {et~al.}(2020){Alfaro}, {Rodriguez}, {Ruiz}, \&
  {Lambas}}]{Alfaro2020}
{Alfaro}, I.~G., {Rodriguez}, F., {Ruiz}, A.~N., \& {Lambas}, D.~G. 2020, \aap,
  638, A60

\bibitem[{{Argudo-Fern{\'a}ndez} {et~al.}(2024){Argudo-Fern{\'a}ndez},
  {G{\'o}mez Hern{\'a}ndez}, {Verley}, {Zurita}, {Duarte Puertas},
  {Bl{\'a}zquez Calero}, {Dom{\'\i}nguez-G{\'o}mez}, {Espada}, {Florido},
  {P{\'e}rez}, \& {S{\'a}nchez-Menguiano}}]{ArgudoFernandez2024}
{Argudo-Fern{\'a}ndez}, M., {G{\'o}mez Hern{\'a}ndez}, C., {Verley}, S.,
  {et~al.} 2024, \aap, 692, A258

\bibitem[{{Bacon} {et~al.}(2001){Bacon}, {Copin}, {Monnet}, {Miller},
  {Allington-Smith}, {Bureau}, {Carollo}, {Davies}, {Emsellem}, {Kuntschner},
  {Peletier}, {Verolme}, \& {de Zeeuw}}]{Bacon01}
{Bacon}, R., {Copin}, Y., {Monnet}, G., {et~al.} 2001, \mnras, 326, 23

\bibitem[{{Barrera-Ballesteros} {et~al.}(2021){Barrera-Ballesteros}, {Heckman},
  {S{\'a}nchez}, {Drory}, {Cruz-Gonzalez}, {Carigi}, {Riffel}, {Boquien},
  {Tissera}, {Bizyaev}, {Rong}, {Boardman}, {Alvarez Hurtado}, \& {MaNGA
  Team}}]{BarreraBallesteros2021a}
{Barrera-Ballesteros}, J.~K., {Heckman}, T., {S{\'a}nchez}, S.~F., {et~al.}
  2021, \apj, 909, 131

\bibitem[{{Brinchmann} {et~al.}(2004){Brinchmann}, {Charlot}, {White},
  {Tremonti}, {Kauffmann}, {Heckman}, \& {Brinkmann}}]{Brinchmann2004}
{Brinchmann}, J., {Charlot}, S., {White}, S.~D.~M., {et~al.} 2004, \mnras, 351,
  1151

\bibitem[{{Bundy} {et~al.}(2015){Bundy}, {Bershady}, {Law}, {Yan}, {Drory},
  {MacDonald}, {Wake}, {Cherinka}, {S{\'a}nchez-Gallego}, {Weijmans}, {Thomas},
  {Tremonti}, {Masters}, {Coccato}, {Diamond-Stanic}, {Arag{\'o}n-Salamanca},
  {Avila-Reese}, {Badenes}, {Falc{\'o}n-Barroso}, {Belfiore}, {Bizyaev},
  {Blanc}, {Bland-Hawthorn}, {Blanton}, {Brownstein}, {Byler}, {Cappellari},
  {Conroy}, {Dutton}, {Emsellem}, {Etherington}, {Frinchaboy}, {Fu}, {Gunn},
  {Harding}, {Johnston}, {Kauffmann}, {Kinemuchi}, {Klaene}, {Knapen},
  {Leauthaud}, {Li}, {Lin}, {Maiolino}, {Malanushenko}, {Malanushenko}, {Mao},
  {Maraston}, {McDermid}, {Merrifield}, {Nichol}, {Oravetz}, {Pan}, {Parejko},
  {Sanchez}, {Schlegel}, {Simmons}, {Steele}, {Steinmetz}, {Thanjavur},
  {Thompson}, {Tinker}, {van den Bosch}, {Westfall}, {Wilkinson}, {Wright},
  {Xiao}, \& {Zhang}}]{Bundy2015}
{Bundy}, K., {Bershady}, M.~A., {Law}, D.~R., {et~al.} 2015, \apj, 798, 7

\bibitem[{{Ceccarelli} {et~al.}(2008){Ceccarelli}, {Padilla}, \&
  {Lambas}}]{Ceccarelli2008}
{Ceccarelli}, L., {Padilla}, N., \& {Lambas}, D.~G. 2008, \mnras, 390, L9

\bibitem[{{Cid Fernandes}(2018)}]{CidFernandes2018}
{Cid Fernandes}, R. 2018, \mnras, 480, 4480

\bibitem[{{Cid Fernandes} {et~al.}(2005){Cid Fernandes}, {Mateus}, {Sodr{\'e}},
  {Stasi{\'n}ska}, \& {Gomes}}]{CidFernandes2005}
{Cid Fernandes}, R., {Mateus}, A., {Sodr{\'e}}, L., {Stasi{\'n}ska}, G., \&
  {Gomes}, J.~M. 2005, \mnras, 358, 363

\bibitem[{{Conrado} {et~al.}(2024){Conrado}, {Gonz{\'a}lez Delgado},
  {Garc{\'\i}a-Benito}, {P{\'e}rez}, {Verley}, {Ruiz-Lara},
  {S{\'a}nchez-Menguiano}, {Duarte Puertas}, {Jim{\'e}nez},
  {Dom{\'\i}nguez-G{\'o}mez}, {Espada}, {Argudo-Fern{\'a}ndez},
  {Alc{\'a}zar-Laynez}, {Bl{\'a}zquez-Calero}, {Bidaran}, {Zurita}, {Peletier},
  {Torres-R{\'\i}os}, {Florido}, {Rodr{\'\i}guez Mart{\'\i}nez}, {del
  Moral-Castro}, {van de Weygaert}, {Falc{\'o}n-Barroso}, {Lugo-Aranda},
  {S{\'a}nchez}, {van der Hulst}, {Courtois}, {Ferr{\'e}-Mateu},
  {S{\'a}nchez-Bl{\'a}zquez}, {Rom{\'a}n}, \& {Aceituno}}]{Conrado2024}
{Conrado}, A.~M., {Gonz{\'a}lez Delgado}, R.~M., {Garc{\'\i}a-Benito}, R.,
  {et~al.} 2024, \aap, 687, A98

\bibitem[{{Croom} {et~al.}(2012){Croom}, {Lawrence}, {Bland-Hawthorn},
  {Bryant}, {Fogarty}, {Richards}, {Goodwin}, {Farrell}, {Miziarski}, {Heald},
  {Jones}, {Lee}, {Colless}, {Brough}, {Hopkins}, {Bauer}, {Birchall}, {Ellis},
  {Horton}, {Leon-Saval}, {Lewis}, {L{\'o}pez-S{\'a}nchez}, {Min}, {Trinh}, \&
  {Trowland}}]{Croom2012MNRAS}
{Croom}, S.~M., {Lawrence}, J.~S., {Bland-Hawthorn}, J., {et~al.} 2012, \mnras,
  421, 872

\bibitem[{{Curtis} {et~al.}(2024){Curtis}, {McDonough}, \&
  {Brainerd}}]{Curtis2024}
{Curtis}, O., {McDonough}, B., \& {Brainerd}, T.~G. 2024, \apj, 962, 58

\bibitem[{{Dey} {et~al.}(2019){Dey}, {Schlegel}, {Lang}, {Blum}, {Burleigh},
  {Fan}, {Findlay}, {Finkbeiner}, {Herrera}, {Juneau}, {Landriau}, {Levi},
  {McGreer}, {Meisner}, {Myers}, {Moustakas}, {Nugent}, {Patej}, {Schlafly},
  {Walker}, {Valdes}, {Weaver}, {Y{\`e}che}, {Zou}, {Zhou}, {Abareshi},
  {Abbott}, {Abolfathi}, {Aguilera}, {Alam}, {Allen}, {Alvarez}, {Annis},
  {Ansarinejad}, {Aubert}, {Beechert}, {Bell}, {BenZvi}, {Beutler}, {Bielby},
  {Bolton}, {Brice{\~n}o}, {Buckley-Geer}, {Butler}, {Calamida}, {Carlberg},
  {Carter}, {Casas}, {Castander}, {Choi}, {Comparat}, {Cukanovaite}, {Delubac},
  {DeVries}, {Dey}, {Dhungana}, {Dickinson}, {Ding}, {Donaldson}, {Duan},
  {Duckworth}, {Eftekharzadeh}, {Eisenstein}, {Etourneau}, {Fagrelius},
  {Farihi}, {Fitzpatrick}, {Font-Ribera}, {Fulmer}, {G{\"a}nsicke},
  {Gaztanaga}, {George}, {Gerdes}, {Gontcho}, {Gorgoni}, {Green}, {Guy},
  {Harmer}, {Hernandez}, {Honscheid}, {Huang}, {James}, {Jannuzi}, {Jiang},
  {Joyce}, {Karcher}, {Karkar}, {Kehoe}, {Kneib}, {Kueter-Young}, {Lan},
  {Lauer}, {Le Guillou}, {Le Van Suu}, {Lee}, {Lesser}, {Perreault Levasseur},
  {Li}, {Mann}, {Marshall}, {Mart{\'\i}nez-V{\'a}zquez}, {Martini}, {du Mas des
  Bourboux}, {McManus}, {Meier}, {M{\'e}nard}, {Metcalfe},
  {Mu{\~n}oz-Guti{\'e}rrez}, {Najita}, {Napier}, {Narayan}, {Newman}, {Nie},
  {Nord}, {Norman}, {Olsen}, {Paat}, {Palanque-Delabrouille}, {Peng},
  {Poppett}, {Poremba}, {Prakash}, {Rabinowitz}, {Raichoor}, {Rezaie},
  {Robertson}, {Roe}, {Ross}, {Ross}, {Rudnick}, {Safonova}, {Saha},
  {S{\'a}nchez}, {Savary}, {Schweiker}, {Scott}, {Seo}, {Shan}, {Silva},
  {Slepian}, {Soto}, {Sprayberry}, {Staten}, {Stillman}, {Stupak}, {Summers},
  {Sien Tie}, {Tirado}, {Vargas-Maga{\~n}a}, {Vivas}, {Wechsler}, {Williams},
  {Yang}, {Yang}, {Yapici}, {Zaritsky}, {Zenteno}, {Zhang}, {Zhang}, {Zhou}, \&
  {Zhou}}]{DeyDESI2019}
{Dey}, A., {Schlegel}, D.~J., {Lang}, D., {et~al.} 2019, \aj, 157, 168

\bibitem[{{Dom{\'\i}nguez-G{\'o}mez} {et~al.}(2022){Dom{\'\i}nguez-G{\'o}mez},
  {Lisenfeld}, {P{\'e}rez}, {L{\'o}pez-S{\'a}nchez}, {Duarte Puertas},
  {Falc{\'o}n-Barroso}, {Kreckel}, {Peletier}, {Ruiz-Lara}, {van de Weygaert},
  {van der Hulst}, \& {Verley}}]{DominguezGomez2022}
{Dom{\'\i}nguez-G{\'o}mez}, J., {Lisenfeld}, U., {P{\'e}rez}, I., {et~al.}
  2022, \aap, 658, A124

\bibitem[{{Dom{\'\i}nguez-G{\'o}mez}
  {et~al.}(2023{\natexlab{a}}){Dom{\'\i}nguez-G{\'o}mez}, {P{\'e}rez},
  {Ruiz-Lara}, {Peletier}, {S{\'a}nchez-Bl{\'a}zquez}, {Lisenfeld}, {Bidaran},
  {Falc{\'o}n-Barroso}, {Alc{\'a}zar-Laynez}, {Argudo-Fern{\'a}ndez},
  {Bl{\'a}zquez-Calero}, {Courtois}, {Duarte Puertas}, {Espada}, {Florido},
  {Garc{\'\i}a-Benito}, {Jim{\'e}nez}, {Kreckel}, {Rela{\~n}o},
  {S{\'a}nchez-Menguiano}, {van der Hulst}, {van de Weygaert}, {Verley}, \&
  {Zurita}}]{Dominguez2023b}
{Dom{\'\i}nguez-G{\'o}mez}, J., {P{\'e}rez}, I., {Ruiz-Lara}, T., {et~al.}
  2023{\natexlab{a}}, \aap, 680, A111

\bibitem[{{Dom{\'\i}nguez-G{\'o}mez}
  {et~al.}(2023{\natexlab{b}}){Dom{\'\i}nguez-G{\'o}mez}, {P{\'e}rez},
  {Ruiz-Lara}, {Peletier}, {S{\'a}nchez-Bl{\'a}zquez}, {Lisenfeld},
  {Falc{\'o}n-Barroso}, {Alc{\'a}zar-Laynez}, {Argudo-Fern{\'a}ndez},
  {Bl{\'a}zquez-Calero}, {Courtois}, {Duarte Puertas}, {Espada}, {Florido},
  {Garc{\'\i}a-Benito}, {Jim{\'e}nez}, {Kreckel}, {Rela{\~n}o},
  {S{\'a}nchez-Menguiano}, {van der Hulst}, {van de Weygaert}, {Verley}, \&
  {Zurita}}]{Dominguez2023a}
{Dom{\'\i}nguez-G{\'o}mez}, J., {P{\'e}rez}, I., {Ruiz-Lara}, T., {et~al.}
  2023{\natexlab{b}}, \nat, 619, 269

\bibitem[{{Douglass} {et~al.}(2019){Douglass}, {Smith}, \&
  {Demina}}]{Douglass2019}
{Douglass}, K.~A., {Smith}, J.~A., \& {Demina}, R. 2019, \apj, 886, 153

\bibitem[{{Florez} {et~al.}(2021){Florez}, {Berlind}, {Kannappan}, {Stark},
  {Eckert}, {Calderon}, {Moffett}, {Campbell}, \& {Sinha}}]{Florez2021}
{Florez}, J., {Berlind}, A.~A., {Kannappan}, S.~J., {et~al.} 2021, \apj, 906,
  97

\bibitem[{{Gallazzi} {et~al.}(2005){Gallazzi}, {Charlot}, {Brinchmann},
  {White}, \& {Tremonti}}]{Gallazzi2005}
{Gallazzi}, A., {Charlot}, S., {Brinchmann}, J., {White}, S. D.~M., \&
  {Tremonti}, C.~A. 2005, \mnras, 362, 41

\bibitem[{{Garc{\'\i}a-Benito} {et~al.}(2017){Garc{\'\i}a-Benito},
  {Gonz{\'a}lez Delgado}, {P{\'e}rez}, {Cid Fernandes}, {Cortijo-Ferrero},
  {L{\'o}pez Fern{\'a}ndez}, {de Amorim}, {Lacerda}, {Vale Asari}, \&
  {S{\'a}nchez}}]{GarciaBenito2017}
{Garc{\'\i}a-Benito}, R., {Gonz{\'a}lez Delgado}, R.~M., {P{\'e}rez}, E.,
  {et~al.} 2017, \aap, 608, A27

\bibitem[{{Gonz{\'a}lez Delgado} {et~al.}(2005){Gonz{\'a}lez Delgado},
  {Cervi{\~n}o}, {Martins}, {Leitherer}, \&
  {Hauschildt}}]{Gonzalez-Delgado2005}
{Gonz{\'a}lez Delgado}, R.~M., {Cervi{\~n}o}, M., {Martins}, L.~P.,
  {Leitherer}, C., \& {Hauschildt}, P.~H. 2005, \mnras, 357, 945

\bibitem[{{Gonz{\'a}lez Delgado} {et~al.}(2014){Gonz{\'a}lez Delgado},
  {P{\'e}rez}, {Cid Fernandes}, {Garc{\'\i}a-Benito}, {de Amorim},
  {S{\'a}nchez}, {Husemann}, {Cortijo-Ferrero}, {L{\'o}pez Fern{\'a}ndez},
  {S{\'a}nchez-Bl{\'a}zquez}, {Bekeraite}, {Walcher}, {Falc{\'o}n-Barroso},
  {Gallazzi}, {van de Ven}, {Alves}, {Bland-Hawthorn}, {Kennicutt}, {Kupko},
  {Lyubenova}, {Mast}, {Moll{\'a}}, {Marino}, {Quirrenbach}, {V{\'\i}lchez}, \&
  {Wisotzki}}]{GonzalezDelgado2014}
{Gonz{\'a}lez Delgado}, R.~M., {P{\'e}rez}, E., {Cid Fernandes}, R., {et~al.}
  2014, \aap, 562, A47

\bibitem[{{Habouzit} {et~al.}(2020){Habouzit}, {Pisani}, {Goulding}, {Dubois},
  {Somerville}, \& {Greene}}]{Habouzit2020}
{Habouzit}, M., {Pisani}, A., {Goulding}, A., {et~al.} 2020, \mnras, 493, 899

\bibitem[{{Hoyle} {et~al.}(2005){Hoyle}, {Rojas}, {Vogeley}, \&
  {Brinkmann}}]{Hoyle2005}
{Hoyle}, F., {Rojas}, R.~R., {Vogeley}, M.~S., \& {Brinkmann}, J. 2005, \apj,
  620, 618

\bibitem[{{Hoyle} {et~al.}(2012){Hoyle}, {Vogeley}, \& {Pan}}]{Hoyle2012}
{Hoyle}, F., {Vogeley}, M.~S., \& {Pan}, D. 2012, \mnras, 426, 3041

\bibitem[{{Ibarra-Medel} {et~al.}(2016){Ibarra-Medel}, {S{\'a}nchez},
  {Avila-Reese}, {Hern{\'a}ndez-Toledo}, {Gonz{\'a}lez}, {Drory}, {Bundy},
  {Bizyaev}, {Cano-D{\'\i}az}, {Malanushenko}, {Pan}, {Roman-Lopes}, \&
  {Thomas}}]{IbarraMedel2016}
{Ibarra-Medel}, H.~J., {S{\'a}nchez}, S.~F., {Avila-Reese}, V., {et~al.} 2016,
  \mnras, 463, 2799

\bibitem[{{Kauffmann} {et~al.}(2003){Kauffmann}, {Heckman}, {White}, {Charlot},
  {Tremonti}, {Brinchmann}, {Bruzual}, {Peng}, {Seibert}, {Bernardi},
  {Blanton}, {Brinkmann}, {Castander}, {Cs{\'a}bai}, {Fukugita}, {Ivezic},
  {Munn}, {Nichol}, {Padmanabhan}, {Thakar}, {Weinberg}, \&
  {York}}]{Kauffman2003}
{Kauffmann}, G., {Heckman}, T.~M., {White}, S. D.~M., {et~al.} 2003, \mnras,
  341, 33

\bibitem[{{Kennicutt}(1998)}]{Kennicutt1998}
{Kennicutt}, Jr., R.~C. 1998, \araa, 36, 189

\bibitem[{{Kreckel} {et~al.}(2011){Kreckel}, {Platen}, {Arag{\'o}n-Calvo}, {van
  Gorkom}, {van de Weygaert}, {van der Hulst}, {Kova{\v{c}}}, {Yip}, \&
  {Peebles}}]{Kreckel2011}
{Kreckel}, K., {Platen}, E., {Arag{\'o}n-Calvo}, M.~A., {et~al.} 2011, \aj,
  141, 4

\bibitem[{{Lacerda} {et~al.}(2022){Lacerda}, {S{\'a}nchez},
  {Mej{\'\i}a-Narv{\'a}ez}, {Camps-Fari{\~n}a}, {Espinosa-Ponce},
  {Barrera-Ballesteros}, {Ibarra-Medel}, \& {Lugo-Aranda}}]{Lacerna2022}
{Lacerda}, E. A.~D., {S{\'a}nchez}, S.~F., {Mej{\'\i}a-Narv{\'a}ez}, A.,
  {et~al.} 2022, \na, 97, 101895

\bibitem[{{Martizzi} {et~al.}(2020){Martizzi}, {Vogelsberger}, {Torrey},
  {Pillepich}, {Hansen}, {Marinacci}, \& {Hernquist}}]{Martizzi2020}
{Martizzi}, D., {Vogelsberger}, M., {Torrey}, P., {et~al.} 2020, \mnras, 491,
  5747

\bibitem[{{Moorman} {et~al.}(2016){Moorman}, {Moreno}, {White}, {Vogeley},
  {Hoyle}, {Giovanelli}, \& {Haynes}}]{Moorman2016}
{Moorman}, C.~M., {Moreno}, J., {White}, A., {et~al.} 2016, \apj, 831, 118

\bibitem[{{Moustakas} {et~al.}(2010){Moustakas}, {Kennicutt}, {Tremonti},
  {Dale}, {Smith}, \& {Calzetti}}]{Moustakas2010}
{Moustakas}, J., {Kennicutt}, Jr., R.~C., {Tremonti}, C.~A., {et~al.} 2010,
  \apjs, 190, 233

\bibitem[{{Nelson} {et~al.}(2019){Nelson}, {Springel}, {Pillepich},
  {Rodriguez-Gomez}, {Torrey}, {Genel}, {Vogelsberger}, {Pakmor}, {Marinacci},
  {Weinberger}, {Kelley}, {Lovell}, {Diemer}, \& {Hernquist}}]{Nelson2019a}
{Nelson}, D., {Springel}, V., {Pillepich}, A., {et~al.} 2019, Computational
  Astrophysics and Cosmology, 6, 2

\bibitem[{{Patiri} {et~al.}(2006){Patiri}, {Prada}, {Holtzman}, {Klypin}, \&
  {Betancort-Rijo}}]{Patiri2006b}
{Patiri}, S.~G., {Prada}, F., {Holtzman}, J., {Klypin}, A., \&
  {Betancort-Rijo}, J. 2006, \mnras, 372, 1710

\bibitem[{{P{\'e}rez} {et~al.}(2013){P{\'e}rez}, {Cid Fernandes}, {Gonz{\'a}lez
  Delgado}, {Garc{\'\i}a-Benito}, {S{\'a}nchez}, {Husemann}, {Mast},
  {Rod{\'o}n}, {Kupko}, {Backsmann}, {de Amorim}, {van de Ven}, {Walcher},
  {Wisotzki}, {Cortijo-Ferrero}, \& {CALIFA Collaboration}}]{PerezE2013}
{P{\'e}rez}, E., {Cid Fernandes}, R., {Gonz{\'a}lez Delgado}, R.~M., {et~al.}
  2013, \apjl, 764, L1

\bibitem[{{P{\'e}rez} {et~al.}(2024){P{\'e}rez}, {Verley},
  {S{\'a}nchez-Menguiano}, {Ruiz-Lara}, {Garc{\'\i}a-Benito}, {Duarte Puertas},
  {Jim{\'e}nez}, {Dom{\'\i}nguez-G{\'o}mez}, {Espada}, {Peletier}, {Rom{\'a}n},
  {Rodr{\'\i}guez}, {Argudo-Fern{\'a}ndez}, {Torres-R{\'\i}os}, {Bidaran},
  {Alc{\'a}zar-Laynez}, {van de Weygaert}, {S{\'a}nchez}, {Lisenfeld},
  {Zurita}, {Florido}, {van der Hulst}, {Bl{\'a}zquez-Calero},
  {Villalba-Gonz{\'a}lez}, {del Moral-Castro}, {S{\'a}nchez Alarc{\'o}n},
  {Lugo-Aranda}, {Walo-Mart{\'\i}n}, {Conrado}, {Gonz{\'a}lez Delgado},
  {Falc{\'o}n-Barroso}, {Ferr{\'e}-Mateu}, {Hern{\'a}ndez-S{\'a}nchez}, {Awad},
  {Kreckel}, {Courtois}, {Espada-Miura}, {Rela{\~n}o}, {Galbany},
  {S{\'a}nchez-Bl{\'a}zquez}, {P{\'e}rez-Montero}, {S{\'a}nchez-Portal},
  {Bongiovanni}, {Planelles}, {Quilis}, {Weijmans}, {Raj}, {Arag{\'o}n-Calvo},
  {Azzaro}, {Bergond}, {Blazek}, {Cikota}, {Fern{\'a}ndez-Mart{\'\i}n},
  {Gardini}, {Guijarro}, {Hermelo}, {Mart{\'\i}n}, \& {Vico
  Linares}}]{Perez2024}
{P{\'e}rez}, I., {Verley}, S., {S{\'a}nchez-Menguiano}, L., {et~al.} 2024,
  \aap, 689, A213

\bibitem[{{Pillepich} {et~al.}(2018){Pillepich}, {Nelson}, {Hernquist},
  {Springel}, {Pakmor}, {Torrey}, {Weinberger}, {Genel}, {Naiman}, {Marinacci},
  \& {Vogelsberger}}]{Pillepich2018}
{Pillepich}, A., {Nelson}, D., {Hernquist}, L., {et~al.} 2018, \mnras, 475, 648

\bibitem[{{Porter} {et~al.}(2023){Porter}, {Holwerda}, {Kruk},
  {Lara-L{\'o}pez}, {Pimbblet}, {Henry}, {Casura}, \& {Kelvin}}]{Porter2023}
{Porter}, L.~E., {Holwerda}, B.~W., {Kruk}, S., {et~al.} 2023, \mnras, 524,
  5768

\bibitem[{{Ricciardelli} {et~al.}(2017){Ricciardelli}, {Cava}, {Varela}, \&
  {Tamone}}]{Ricciardelli2017}
{Ricciardelli}, E., {Cava}, A., {Varela}, J., \& {Tamone}, A. 2017, \apjl, 846,
  L4

\bibitem[{{Riffel} {et~al.}(2023){Riffel}, {Mallmann}, {Rembold}, {Ilha},
  {Riffel}, {Storchi-Bergmann}, {Ruschel-Dutra}, {Vazdekis},
  {Mart{\'\i}n-Navarro}, {Schimoia}, {Ramos Almeida}, {da Costa}, {Vila-Verde},
  \& {Gatto}}]{Riffel2023}
{Riffel}, R., {Mallmann}, N.~D., {Rembold}, S.~B., {et~al.} 2023, \mnras, 524,
  5640

\bibitem[{{Rodriguez} \& {Merch{\'a}n}(2020)}]{Rodriguez2020}
{Rodriguez}, F. \& {Merch{\'a}n}, M. 2020, \aap, 636, A61

\bibitem[{{Rodr{\'\i}guez-Medrano} {et~al.}(2023){Rodr{\'\i}guez-Medrano},
  {Paz}, {Stasyszyn}, {Rodr{\'\i}guez}, {Ruiz}, \&
  {Merch{\'a}n}}]{RodriguezMedrano2023}
{Rodr{\'\i}guez-Medrano}, A.~M., {Paz}, D.~J., {Stasyszyn}, F.~A., {et~al.}
  2023, \mnras, 521, 916

\bibitem[{{Rodr{\'\i}guez Medrano} {et~al.}(2022){Rodr{\'\i}guez Medrano},
  {Paz}, {Stasyszyn}, \& {Ruiz}}]{RodriguezMedrano2022}
{Rodr{\'\i}guez Medrano}, A.~M., {Paz}, D.~J., {Stasyszyn}, F.~A., \& {Ruiz},
  A.~N. 2022, \mnras, 511, 2688

\bibitem[{{Rodr{\'\i}guez-Medrano} {et~al.}(2024){Rodr{\'\i}guez-Medrano},
  {Springel}, {Stasyszyn}, \& {Paz}}]{RodriguezMedrano2024}
{Rodr{\'\i}guez-Medrano}, A.~M., {Springel}, V., {Stasyszyn}, F.~A., \& {Paz},
  D.~J. 2024, \mnras, 528, 2822

\bibitem[{{Rojas} {et~al.}(2004){Rojas}, {Vogeley}, {Hoyle}, \&
  {Brinkmann}}]{Rojas2004}
{Rojas}, R.~R., {Vogeley}, M.~S., {Hoyle}, F., \& {Brinkmann}, J. 2004, \apj,
  617, 50

\bibitem[{{Rojas} {et~al.}(2005){Rojas}, {Vogeley}, {Hoyle}, \&
  {Brinkmann}}]{Rojas2005}
{Rojas}, R.~R., {Vogeley}, M.~S., {Hoyle}, F., \& {Brinkmann}, J. 2005, \apj,
  624, 571

\bibitem[{{Rosas-Guevara} {et~al.}(2022){Rosas-Guevara}, {Tissera}, {Lagos},
  {Paillas}, \& {Padilla}}]{RosasGuevara2022}
{Rosas-Guevara}, Y., {Tissera}, P., {Lagos}, C. d.~P., {Paillas}, E., \&
  {Padilla}, N. 2022, \mnras [\eprint[arXiv]{2204.04565}]

\bibitem[{{Ruiz} {et~al.}(2019){Ruiz}, {Alfaro}, \& {Garcia Lambas}}]{Ruiz2019}
{Ruiz}, A.~N., {Alfaro}, I.~G., \& {Garcia Lambas}, D. 2019, \mnras, 483, 4070

\bibitem[{{Ruiz} {et~al.}(2015){Ruiz}, {Paz}, {Lares}, {Luparello},
  {Ceccarelli}, \& {Lambas}}]{Ruiz2015}
{Ruiz}, A.~N., {Paz}, D.~J., {Lares}, M., {et~al.} 2015, \mnras, 448, 1471

\bibitem[{Ruschel-Dutra \& Dall'Agnol De~Oliveira(2020)}]{Ruschel-Dutra2020}
Ruschel-Dutra, D. \& Dall'Agnol De~Oliveira, B. 2020, danielrd6/ifscube:
  Modeling, \url{https://zenodo.org/record/4065550}, dOI:
  \href{https://doi.org/10.5281/zenodo.4065550}{10.5281/zenodo.4065550}

\bibitem[{{Ruschel-Dutra} {et~al.}(2021){Ruschel-Dutra}, {Storchi-Bergmann},
  {Schnorr-M{\"u}ller}, {Riffel}, {Dall'Agnol de Oliveira}, {Lena}, {Robinson},
  {Nagar}, \& {Elvis}}]{Ruschel-Dutra2021}
{Ruschel-Dutra}, D., {Storchi-Bergmann}, T., {Schnorr-M{\"u}ller}, A., {et~al.}
  2021, \mnras, 507, 74

\bibitem[{{S{\'a}nchez} {et~al.}(2022){S{\'a}nchez}, {Barrera-Ballesteros},
  {Lacerda}, {Mej{\'\i}a-Narvaez}, {Camps-Fari{\~n}a}, {Bruzual},
  {Espinosa-Ponce}, {Rodr{\'\i}guez-Puebla}, {Calette}, {Ibarra-Medel},
  {Avila-Reese}, {Hernandez-Toledo}, {Bershady}, {Cano-Diaz}, \&
  {Munguia-Cordova}}]{Sanchez2022}
{S{\'a}nchez}, S.~F., {Barrera-Ballesteros}, J.~K., {Lacerda}, E., {et~al.}
  2022, \apjs, 262, 36

\bibitem[{{S{\'a}nchez} {et~al.}(2024){S{\'a}nchez}, {Garc{\'\i}a-Benito},
  {Gonz{\'a}lez Delgado}, {Conrado}, {Perez}, {Lugo-Aranda},
  {S{\'a}nchez-Menguiano}, {Ruiz-Lara}, {Jim{\'e}nez}, {Duarte Puertas},
  {Dom{\'\i}nguez-G{\'o}mez}, {Torres-R{\'\i}os}, {Argudo-Fern{\'a}ndez},
  {Bl{\'a}zquez-Calero}, {Alc{\'a}zar-Laynez}, {Verley}, {Espada}, {Lisenfeld},
  {Zurita}, {Florido}, {Bidaran}, {Villalba-Gonz{\'a}lez}, {Ferr{\'e}-Mateu},
  {S{\'a}nchez Alarc{\'o}n}, {Rom{\'a}n}, {del Moral-Castro}, \&
  {Ag{\"u}i}}]{Sanchez2024}
{S{\'a}nchez}, S.~F., {Garc{\'\i}a-Benito}, R., {Gonz{\'a}lez Delgado}, R.,
  {et~al.} 2024, \rmxaa, 60, 323

\bibitem[{{S{\'a}nchez} {et~al.}(2012){S{\'a}nchez}, {Kennicutt}, {Gil de Paz},
  {van de Ven}, {V{\'\i}lchez}, {Wisotzki}, {Walcher}, {Mast}, {Aguerri},
  {Albiol-P{\'e}rez}, {Alonso-Herrero}, {Alves}, {Bakos}, {Bart{\'a}kov{\'a}},
  {Bland-Hawthorn}, {Boselli}, {Bomans}, {Castillo-Morales}, {Cortijo-Ferrero},
  {de Lorenzo-C{\'a}ceres}, {Del Olmo}, {Dettmar}, {D{\'\i}az}, {Ellis},
  {Falc{\'o}n-Barroso}, {Flores}, {Gallazzi}, {Garc{\'\i}a-Lorenzo},
  {Gonz{\'a}lez Delgado}, {Gruel}, {Haines}, {Hao}, {Husemann},
  {Igl{\'e}sias-P{\'a}ramo}, {Jahnke}, {Johnson}, {Jungwiert}, {Kalinova},
  {Kehrig}, {Kupko}, {L{\'o}pez-S{\'a}nchez}, {Lyubenova}, {Marino},
  {M{\'a}rmol-Queralt{\'o}}, {M{\'a}rquez}, {Masegosa}, {Meidt},
  {Mendez-Abreu}, {Monreal-Ibero}, {Montijo}, {Mour{\~a}o}, {Palacios-Navarro},
  {Papaderos}, {Pasquali}, {Peletier}, {P{\'e}rez}, {P{\'e}rez}, {Quirrenbach},
  {Rela{\~n}o}, {Rosales-Ortega}, {Roth}, {Ruiz-Lara},
  {S{\'a}nchez-Bl{\'a}zquez}, {Sengupta}, {Singh}, {Stanishev}, {Trager},
  {Vazdekis}, {Viironen}, {Wild}, {Zibetti}, \& {Ziegler}}]{Sanchez2012AA}
{S{\'a}nchez}, S.~F., {Kennicutt}, R.~C., {Gil de Paz}, A., {et~al.} 2012,
  \aap, 538, A8

\bibitem[{{S{\'a}nchez} {et~al.}(2016{\natexlab{a}}){S{\'a}nchez}, {P{\'e}rez},
  {S{\'a}nchez-Bl{\'a}zquez}, {Garc{\'\i}a-Benito}, {Ibarra-Mede},
  {Gonz{\'a}lez}, {Rosales-Ortega}, {S{\'a}nchez-Menguiano}, {Ascasibar},
  {Bitsakis}, {Law}, {Cano-D{\'\i}az}, {L{\'o}pez-Cob{\'a}}, {Marino}, {Gil de
  Paz}, {L{\'o}pez-S{\'a}nchez}, {Barrera-Ballesteros}, {Galbany}, {Mast},
  {Abril-Melgarejo}, \& {Roman-Lopes}}]{Sanchez2016bRMxAA}
{S{\'a}nchez}, S.~F., {P{\'e}rez}, E., {S{\'a}nchez-Bl{\'a}zquez}, P., {et~al.}
  2016{\natexlab{a}}, \rmxaa, 52, 171

\bibitem[{{S{\'a}nchez} {et~al.}(2016{\natexlab{b}}){S{\'a}nchez}, {P{\'e}rez},
  {S{\'a}nchez-Bl{\'a}zquez}, {Gonz{\'a}lez}, {Ros{\'a}les-Ortega},
  {Cano-D{\'\i}az}, {L{\'o}pez-Cob{\'a}}, {Marino}, {Gil de Paz}, {Moll{\'a}},
  {L{\'o}pez-S{\'a}nchez}, {Ascasibar}, \&
  {Barrera-Ballesteros}}]{Sanchez2016aRMxAA}
{S{\'a}nchez}, S.~F., {P{\'e}rez}, E., {S{\'a}nchez-Bl{\'a}zquez}, P., {et~al.}
  2016{\natexlab{b}}, \rmxaa, 52, 21

\bibitem[{{Scholz-D{\'\i}az} {et~al.}(2022){Scholz-D{\'\i}az},
  {Mart{\'\i}n-Navarro}, \& {Falc{\'o}n-Barroso}}]{ScholtzDiaz2022}
{Scholz-D{\'\i}az}, L., {Mart{\'\i}n-Navarro}, I., \& {Falc{\'o}n-Barroso}, J.
  2022, \mnras, 511, 4900

\bibitem[{{Scholz-D{\'\i}az} {et~al.}(2023){Scholz-D{\'\i}az},
  {Mart{\'\i}n-Navarro}, \& {Falc{\'o}n-Barroso}}]{ScholtDiaz2023}
{Scholz-D{\'\i}az}, L., {Mart{\'\i}n-Navarro}, I., \& {Falc{\'o}n-Barroso}, J.
  2023, \mnras, 518, 6325

\bibitem[{{Springel} {et~al.}(2018){Springel}, {Pakmor}, {Pillepich},
  {Weinberger}, {Nelson}, {Hernquist}, {Vogelsberger}, {Genel}, {Torrey},
  {Marinacci}, \& {Naiman}}]{Springel2018}
{Springel}, V., {Pakmor}, R., {Pillepich}, A., {et~al.} 2018, \mnras, 475, 676

\bibitem[{{Tavasoli} {et~al.}(2015){Tavasoli}, {Rahmani}, {Khosroshahi},
  {Vasei}, \& {Lehnert}}]{Tavasoli2015}
{Tavasoli}, S., {Rahmani}, H., {Khosroshahi}, H.~G., {Vasei}, K., \& {Lehnert},
  M.~D. 2015, \apjl, 803, L13

\bibitem[{{Tonnesen} \& {Cen}(2015)}]{Tonnensen2015}
{Tonnesen}, S. \& {Cen}, R. 2015, \apj, 812, 104

\bibitem[{{Torres-R{\'\i}os} {et~al.}(2024){Torres-R{\'\i}os}, {P{\'e}rez},
  {Verley}, {Dom{\'\i}nguez-G{\'o}mez}, {Argudo-Fern{\'a}ndez}, {Duarte
  Puertas}, {Jim{\'e}nez}, {Ruiz-Lara}, {Zurita}, {Bidaran}, {Conrado},
  {Espada}, {Garc{\'\i}a-Benito}, {Gonz{\'a}lez Delgado}, {Falc{\'o}n-Barroso},
  {Florido}, {S{\'a}nchez-Bl{\'a}zquez}, \&
  {S{\'a}nchez-Menguiano}}]{Torres-Rios2024}
{Torres-R{\'\i}os}, G., {P{\'e}rez}, I., {Verley}, S., {et~al.} 2024, \aap,
  691, A341

\bibitem[{{van de Weygaert} \& {Platen}(2011)}]{VanDeWeygaert2011}
{van de Weygaert}, R. \& {Platen}, E. 2011, in International Journal of Modern
  Physics Conference Series, Vol.~1, International Journal of Modern Physics
  Conference Series, 41--66

\bibitem[{{Vazdekis} {et~al.}(2016){Vazdekis}, {Koleva}, {Ricciardelli},
  {R{\"o}ck}, \& {Falc{\'o}n-Barroso}}]{Vazdekis2016}
{Vazdekis}, A., {Koleva}, M., {Ricciardelli}, E., {R{\"o}ck}, B., \&
  {Falc{\'o}n-Barroso}, J. 2016, \mnras, 463, 3409

\bibitem[{{Vazdekis} {et~al.}(2010){Vazdekis}, {S{\'a}nchez-Bl{\'a}zquez},
  {Falc{\'o}n-Barroso}, {Cenarro}, {Beasley}, {Cardiel}, {Gorgas}, \&
  {Peletier}}]{Vazdekis2010}
{Vazdekis}, A., {S{\'a}nchez-Bl{\'a}zquez}, P., {Falc{\'o}n-Barroso}, J.,
  {et~al.} 2010, \mnras, 404, 1639

\bibitem[{{V{\'a}zquez-Mata} {et~al.}(2022){V{\'a}zquez-Mata},
  {Hern{\'a}ndez-Toledo}, {Avila-Reese}, {Herrera-Endoqui},
  {Rodr{\'\i}guez-Puebla}, {Cano-D{\'\i}az}, {Lacerna},
  {Mart{\'\i}nez-V{\'a}zquez}, \& {Lane}}]{Vazquez-Mata2022}
{V{\'a}zquez-Mata}, J.~A., {Hern{\'a}ndez-Toledo}, H.~M., {Avila-Reese}, V.,
  {et~al.} 2022, \mnras, 512, 2222

\bibitem[{{von Benda-Beckmann} \& {M{\"u}ller}(2008)}]{vonBenda2008}
{von Benda-Beckmann}, A.~M. \& {M{\"u}ller}, V. 2008, \mnras, 384, 1189

\bibitem[{{Wake} {et~al.}(2017){Wake}, {Bundy}, {Diamond-Stanic}, {Yan},
  {Blanton}, {Bershady}, {S{\'a}nchez-Gallego}, {Drory}, {Jones}, {Kauffmann},
  {Law}, {Li}, {MacDonald}, {Masters}, {Thomas}, {Tinker}, {Weijmans}, \&
  {Brownstein}}]{Wake2017}
{Wake}, D.~A., {Bundy}, K., {Diamond-Stanic}, A.~M., {et~al.} 2017, \aj, 154,
  86

\bibitem[{{Yan} {et~al.}(2016){Yan}, {Bundy}, {Law}, {Bershady}, {Andrews},
  {Cherinka}, {Diamond-Stanic}, {Drory}, {MacDonald}, {S{\'a}nchez-Gallego},
  {Thomas}, {Wake}, {Weijmans}, {Westfall}, {Zhang}, {Arag{\'o}n-Salamanca},
  {Belfiore}, {Bizyaev}, {Blanc}, {Blanton}, {Brownstein}, {Cappellari},
  {D'Souza}, {Emsellem}, {Fu}, {Gaulme}, {Graham}, {Goddard}, {Gunn},
  {Harding}, {Jones}, {Kinemuchi}, {Li}, {Li}, {Maiolino}, {Mao}, {Maraston},
  {Masters}, {Merrifield}, {Oravetz}, {Pan}, {Parejko}, {Sanchez}, {Schlegel},
  {Simmons}, {Thanjavur}, {Tinker}, {Tremonti}, {van den Bosch}, \&
  {Zheng}}]{Yan2016}
{Yan}, R., {Bundy}, K., {Law}, D.~R., {et~al.} 2016, \aj, 152, 197

\bibitem[{{Yan} {et~al.}(2019){Yan}, {Chen}, {Lazarz}, {Bizyaev}, {Maraston},
  {Stringfellow}, {McCarthy}, {Meneses-Goytia}, {Law}, {Thomas}, {Falcon
  Barroso}, {S{\'a}nchez-Gallego}, {Schlafly}, {Zheng}, {Argudo-Fern{\'a}ndez},
  {Beaton}, {Beers}, {Bershady}, {Blanton}, {Brownstein}, {Bundy}, {Chambers},
  {Cherinka}, {De Lee}, {Drory}, {Galbany}, {Holtzman}, {Imig}, {Kaiser},
  {Kinemuchi}, {Liu}, {Luo}, {Magnier}, {Majewski}, {Nair}, {Oravetz},
  {Oravetz}, {Pan}, {Sobeck}, {Stassun}, {Talbot}, {Tremonti}, {Waters},
  {Weijmans}, {Wilhelm}, {Zasowski}, {Zhao}, \& {Zhao}}]{Yan2019}
{Yan}, R., {Chen}, Y., {Lazarz}, D., {et~al.} 2019, \apj, 883, 175

\bibitem[{{York} {et~al.}(2000){York}, {Adelman}, {Anderson}, {Anderson},
  {Annis}, {Bahcall}, {Bakken}, {Barkhouser}, {Bastian}, {Berman}, {Boroski},
  {Bracker}, {Briegel}, {Briggs}, {Brinkmann}, {Brunner}, {Burles}, {Carey},
  {Carr}, {Castander}, {Chen}, {Colestock}, {Connolly}, {Crocker}, {Csabai},
  {Czarapata}, {Davis}, {Doi}, {Dombeck}, {Eisenstein}, {Ellman}, {Elms},
  {Evans}, {Fan}, {Federwitz}, {Fiscelli}, {Friedman}, {Frieman}, {Fukugita},
  {Gillespie}, {Gunn}, {Gurbani}, {de Haas}, {Haldeman}, {Harris}, {Hayes},
  {Heckman}, {Hennessy}, {Hindsley}, {Holm}, {Holmgren}, {Huang}, {Hull},
  {Husby}, {Ichikawa}, {Ichikawa}, {Ivezi{\'c}}, {Kent}, {Kim}, {Kinney},
  {Klaene}, {Kleinman}, {Kleinman}, {Knapp}, {Korienek}, {Kron}, {Kunszt},
  {Lamb}, {Lee}, {Leger}, {Limmongkol}, {Lindenmeyer}, {Long}, {Loomis},
  {Loveday}, {Lucinio}, {Lupton}, {MacKinnon}, {Mannery}, {Mantsch}, {Margon},
  {McGehee}, {McKay}, {Meiksin}, {Merelli}, {Monet}, {Munn}, {Narayanan},
  {Nash}, {Neilsen}, {Neswold}, {Newberg}, {Nichol}, {Nicinski}, {Nonino},
  {Okada}, {Okamura}, {Ostriker}, {Owen}, {Pauls}, {Peoples}, {Peterson},
  {Petravick}, {Pier}, {Pope}, {Pordes}, {Prosapio}, {Rechenmacher}, {Quinn},
  {Richards}, {Richmond}, {Rivetta}, {Rockosi}, {Ruthmansdorfer}, {Sand ford},
  {Schlegel}, {Schneider}, {Sekiguchi}, {Sergey}, {Shimasaku}, {Siegmund},
  {Smee}, {Smith}, {Snedden}, {Stone}, {Stoughton}, {Strauss}, {Stubbs},
  {SubbaRao}, {Szalay}, {Szapudi}, {Szokoly}, {Thakar}, {Tremonti}, {Tucker},
  {Uomoto}, {Vanden Berk}, {Vogeley}, {Waddell}, {Wang}, {Watanabe},
  {Weinberg}, {Yanny}, {Yasuda}, \& {SDSS Collaboration}}]{York2000}
{York}, D.~G., {Adelman}, J., {Anderson}, John~E., J., {et~al.} 2000, \aj, 120,
  1579

\end{thebibliography}

\begin{appendix}

\section{Samples characterization}
\label{sec:Samples-characterization}
\begin{figure}
	\includegraphics[width=\linewidth]{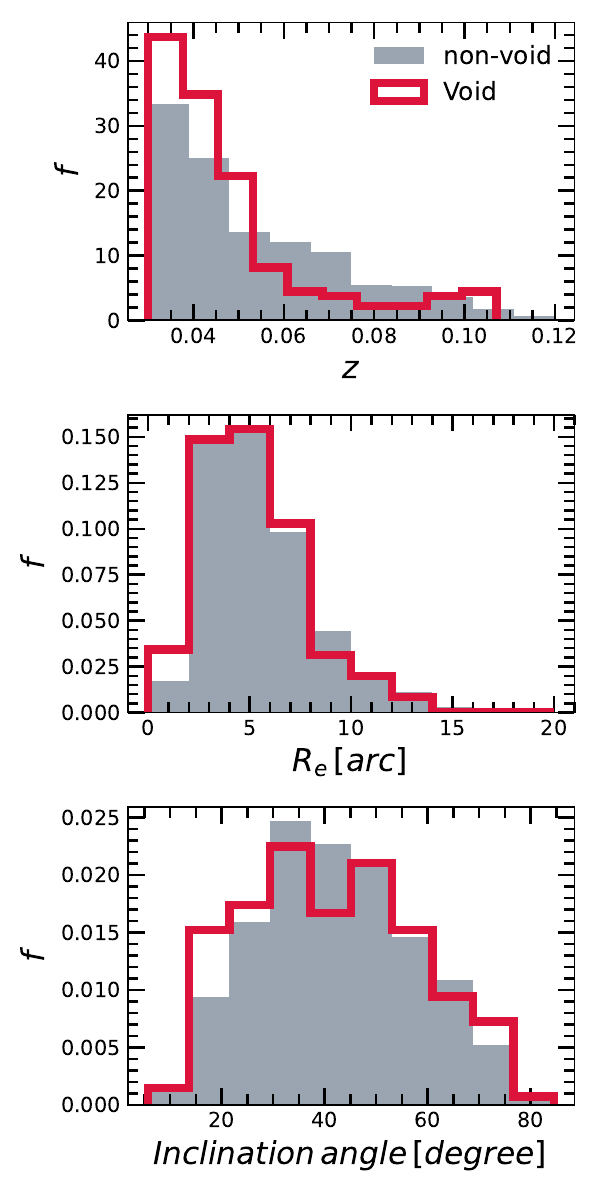}
    \caption{ Properties of void and non-void sample of galaxies. \textit{Top panel:} redshift distribution, \textit{Central panel:} effective radius $R_e\,[arc]$, \textit{Bottom panel:} inclination angle. The grey distribution correspond to non-void galaxies and red to void galaxies.  
    }
    \label{fig:z-Re-inc-distributions}
\end{figure}

\begin{figure}
	\includegraphics[width=\linewidth]{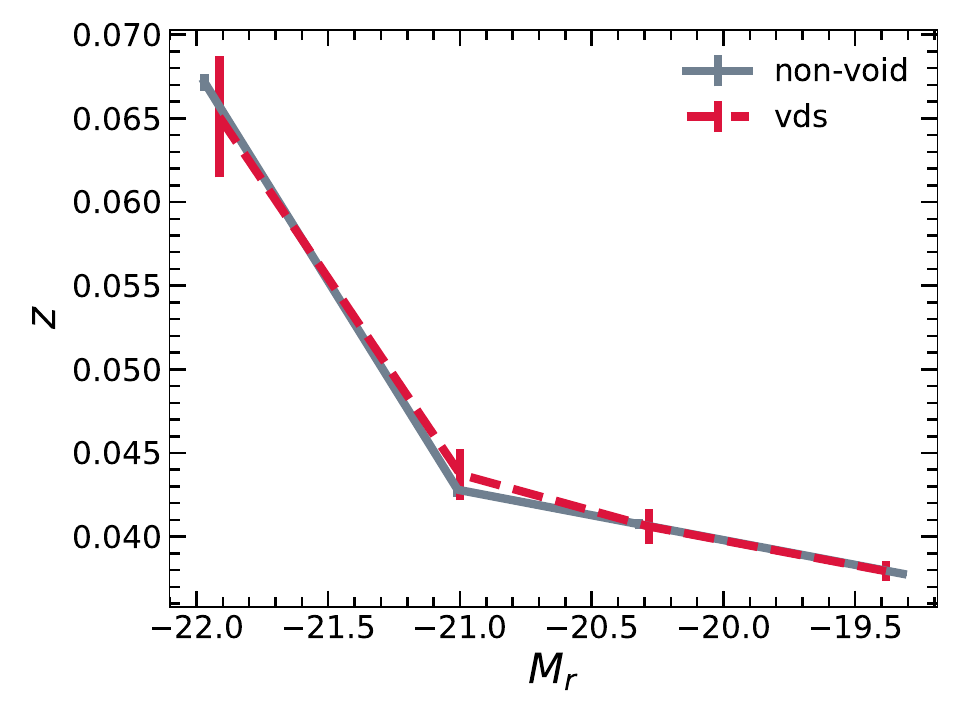}
    \caption{ Mean redshift value as a function of r-band absolute magnitude. The dashed-red line indicate void galaxies and solid grey line correspond to the non-void sample. The error band indicate the error in the mean value. 
    }
    \label{fig:z-mag}
\end{figure}

In Fig. \ref{fig:z-Re-inc-distributions}, we present the distributions of redshift, effective radius, and inclination angle for our sample of galaxies (void and non-void). The figure shows similar distributions for the effective radius and inclination angle. The largest difference is observed in the redshift distribution. This difference arises from the variation in the magnitude of galaxies in both environments (see Fig. \ref{fig:magnitude_dist}). In void environments, the galaxy distribution is biased towards faint galaxies, which cannot be observed at higher redshifts.

To ensure that the difference in the redshift distribution does not bias our results when comparing void and non-void galaxies, we show in Fig. \ref{fig:z-mag} the mean redshift as a function of r-band magnitude for both environments. This plot is analogous to those shown in Fig. \ref{fig:color-sfr} and Fig. \ref{fig:met-age-gas}. From this, we can see that the mean redshift in all magnitude bins is comparable, ensuring that no bias is introduced by differences in the redshift of galaxies.

\begin{figure}
	\includegraphics[width=\linewidth]{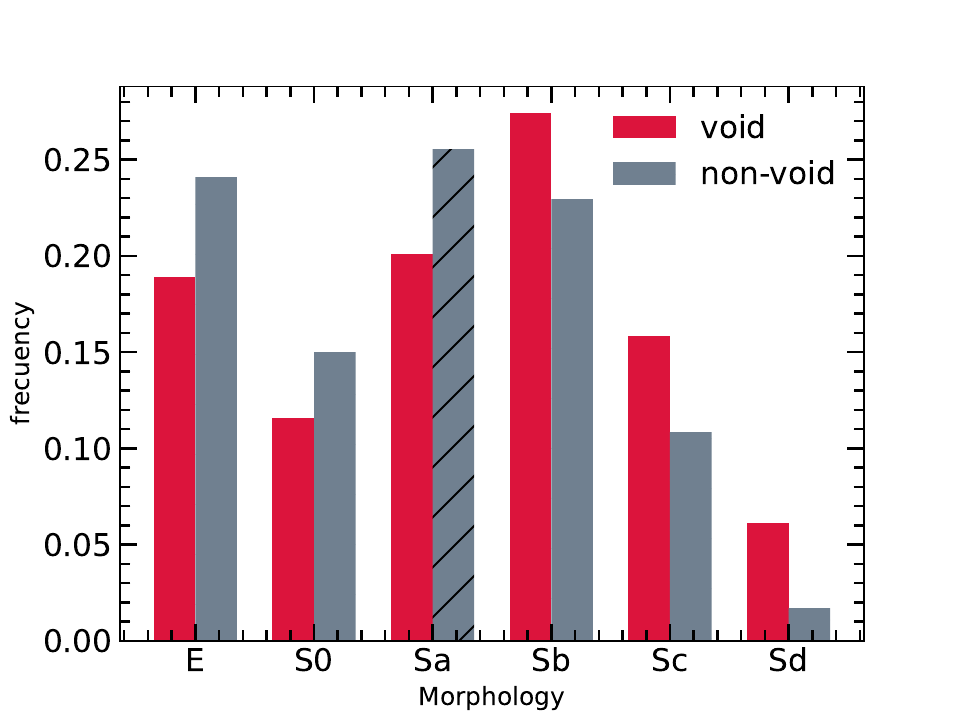}
    \caption{ Distribution of morphological types (E, S0, Sa, Sb, Sc, Sd) for galaxies. With red bars we show the void galaxy sample and with dashed grey bars we show the non-void sample. 
    }
    \label{fig:bar_morphology_completed}
\end{figure}

Figure \ref{fig:bar_morphology_completed} shows the complete morphological type distribution of our samples. Irregular and Sm galaxies were excluded from the plot, as none belong to void environments. The figure illustrates that early-type galaxies (E and S0), as well as Sa, are more abundant in the non-void sample compared to void environments. In contrast, later-type galaxies (Sb, Sc, Sd) are more prevalent in the void sample.

\end{appendix}

\end{document}